\documentclass[12pt]{article}

\usepackage{epsfig}
\usepackage{epstopdf}
\usepackage{fp}
\usepackage{color}
\usepackage{graphicx}
\usepackage[toc,page]{appendix}
\usepackage{natbib}
\usepackage{multicol}
\usepackage[margin=3cm]{geometry}
\graphicspath{ {figs/} }

\newcommand{\Rs}{$ R_{\odot}$}

\newcommand{\pB}{\textit{pB}}

\newcommand{\Bk}{$B_{k}$}

\begin{document}

\title{Erupting filaments with large enclosing flux tubes as sources of high-mass 3-part CMEs, and erupting filaments in the absence of enclosing flux tubes as sources of low-mass unstructured CMEs}
\author{Joe Hutton and Huw Morgan}
\date{\small{Published in The Astrophysical Journal, 813:35, 2015 November 1}}
\maketitle
\begin{center}
\small{Institute of Mathematics, Physics $\&$ Computer Sciences, Aberystwyth University, Penglais, Aberystwyth, Ceredigion, SY23 3BZ}\\ 
\small{joh9@aber.ac.uk}
\end{center}

\begin{abstract}
The 3-part appearance of many CMEs arising from erupting filaments emerges from a large magnetic flux tube structure, consistent with the form of the erupting filament system. Other CMEs arising from erupting filaments lack a clear 3-part structure and reasons for this have not been researched in detail. This paper aims to further establish the link between CME structure and the structure of the erupting filament system and to investigate whether CMEs which lack a 3-part structure have different eruption characteristics. A survey is made of 221 near-limb filament eruptions observed from 2013/05/03-2014/06/30 by EUV imagers and coronagraphs. 92 filament eruptions are associated with 3-part structured CMEs, 41 eruptions are associated with unstructured CMEs. The remaining 88 are categorized as failed eruptions. For 34\% of the 3-part CMEs, processing applied to EUV images reveals the erupting front edge is a pre-existing loop structure surrounding the filament, which subsequently erupts with the filament to form the leading bright front edge of the CME. This connection is confirmed by a flux-rope density model. Furthermore, the unstructured CMEs have a narrower distribution of mass compared to structured CMEs, with  total mass comparable to the mass of 3-part CME cores. This study supports the interpretation of 3-part CME leading fronts as the outer boundaries of a large pre-existing flux tube. Unstructured (non 3-part) CMEs are a different family to structured CMEs, arising from the eruption of filaments which are compact flux tubes in the absence of a large system of enclosing closed field.
\end{abstract}
\small{\textit{keywords} - Sun: corona---Sun: CMEs---Sun:filaments}

%Intro
\section{Introduction}
\label{sec:intro}
Observed by white light coronagraphs in the extended corona, Coronal Mass Ejections (CMEs) are huge eruptions of magnetised plasma which possess a broad range of masses and outflow speeds \citep{lui10,yashiro04}. These eruptions and their associated bursts of energetic particles can cause adverse space weather at Earth, such as geomagnetic storms, which can lead to the damage of satellites, power grid failures and the disruption of communications and GPS \citep{schwenn05}. Due to these potential risks, the physics governing their eruption and propagation through the heliosphere needs to be understood.

The eruptions of solar filaments and their associated cavities (and/or overlaying arcade systems) are a major source of CMEs. A statistical study by \citet{jing04} showed that 56$\%$ of filament eruptions were associated with CMEs. The dense material of the erupting filament is then later observed as the bright inner cores of CMEs as described in the 3-part structure model: the bright CME leading front, followed by a cavity and a bright central core \citep{vourlidas2013,chen11}. There is a strong connection therefore between general studies and models of filaments and their associated cavities and the CMEs observed when they erupt \citep[e.g.][]{cheng2014}. In a quiescent state in the low corona, the filament, or filament-cavity system is a magnetic flux tube. The dense filament exists due to the larger surrounding flux tube \citep{low2012a,berger2012}. The so-called cavity contains low-density hot coronal plasma which, entrapped in the flux tube, can condensate and sink, forming a high-density region in the lowest part of the flux tube. This region has a lower temperature than the surrounding cavity system due to thin radiative cooling \citep{gibson2010, habbal2010, liu2012}. If a flux tube in an equilibrium state is elevated in the corona, this material can pass through the lower part of the flux tube and collect in the complex field surrounding the underlying current sheet forming spectacular prominences above the solar limb \citep{low2012b}.

The appearance of the filament-cavity flux tube system can vary greatly according to the relative alignment of the flux tube axis with the observer \citep{habbal2014}, or if it is observed against the disk or above the limb. The quiescent filament-cavity flux tube is anchored to the Sun by overlying magnetic field arcade and by the field of the flux tube itself linked to the photosphere. Eruptions occur due to instabilities formed during the evolution of the external magnetic field \citep{chen11} or the field of the flux tube itself. The processes causing this can be tether-cutting (a weakening of the field connecting the flux rope to the Sun by reconnection, \citet{moore92}), tether-cutting caused by emerging flux near the base of the filament system \citep{chen00,sterling07}, breakout (a weakening of the overlying arcade field, \citet{antiochos99}), or an instability of the flux tube due to twisting \citep[e.g.][]{torok05}. Recent observations show that filament-cavity systems may unravel without eruption by reconnection with neighbouring open field, leading to an outward-propagating twist in the corona's open field \citep{habbal2014}. Active region filaments are smaller and shorter-lived, embedded within active regions above the polarity inversion line (see \citet{mackay2010,labrosse2010} for an extensive review). They exist within the surrounding closed field which is often complex and dynamic, and the concept of a surrounding cavity (or enclosing flux tube structure) is not so applicable, and certainly difficult to observe.

Structured 3-part CMEs observed in white light in the extended corona, are well-modeled as large flux tubes, with the CME leading front as the front boundary of the tube, expanding and broadening in the background coronal plasma and magnetic field. The bright inner core is formed from the dense material of the erupted filament. Several studies have found good agreement between the observed structure of white light CMEs and a model wire-frame of a flux-tube: for example, \citet{thernisien09} describes the resulting shape of a CME as being ``reminiscent of a hollow croissant''. UltraViolet spectroscopy of CMEs show emission from cool material in CME cores, thus giving direct evidence of their source as dense filaments \citep{landi2010}. Many {\em in situ} magnetic field measurements of Interplanetary CMEs (ICMEs) agree well with a flux rope model \citep[e.g.][]{gop2006}, with direct measurement of the filament material \citep{lepri2010}. The model of erupting filaments and CMEs as flux tubes is well-established to the point that some scientists even pose the challenge of identifying CMEs that are clearly not flux rope CMEs (FR-CMEs) \citep{vourlidas2013}. As an alternative, some suggest that CME leading fronts are wave phenomena, that is, fast-mode MHD waves excited by the pressure pulse from the underlying erupting filament \citep{chen11}. Providing a link between these two possible sources is clear observational evidence of a 5-part structure to some white-light CMEs. This is the standard 3-part structure, but with a shock and depletion region leading the bright front-edge of the CME \citep{jackson1978,vourlidas2013} - thus there is a faint yet observable shock region leading the front structure of the 3-part CME. %A convincing explanation has been made of the general differing appearance of CME 3-part structure in the East and West corona according to the relative alignment of the longitudinally-aligned filaments prior to eruptions \citep{***wishwerememberedthispaper}.

The main source of uncertainty in linking the structure of erupting filaments, observed in the lowest corona, and the structure of the resulting white-light CMEs, is the current observational limitations due to the lack of quality coronagraph observations below a height of $\sim$1\Rs\ above the limb. Extreme UltraViolet (EUV) imaging of the lowest corona has become routine, giving the main source of information on the initial eruptions of CMEs. The EUV signal, however, drops very quickly with height, effectively restricting current observations to below $\sim0.3$\Rs\ above the limb. The collisional excitation of coronal EUV lines leads to a drop in intensity proportional to density squared, leading to an inherent rapid drop in signal with height \citep[e.g.][]{habbal2013}. There is therefore a cruicial gap in observation between $\sim1.3-2.2$\Rs. Above $\sim$2.2\Rs, the Large Angle Spectrometric Coronagraph (LASCO) C2 coronagraph gives quality observations of the corona in broadband visible light. The Sun Watcher using Active Pixel System Detector and Image Processing (SWAP) instrument aboard the Projects for Onboard Autonomy (PROBA2) is an EUV imager which can observe to larger heights given specific observational sequences \citep{seaton2013,halain2013}. 

This study shows how structural connections between the erupting filaments and white light CMEs may be made by the use of a flux-rope model, with the position and propagation of the flux rope constrained by the kinematics of the observed filament and CME. Recently developed image processing techniques are also employed, which aid in both revealing important structural detail, including the CME front edge, in EUV images \citep{mgn}, and in white light coronagraph images \citep{auto1}. The instruments and methods are presented in section \ref{sec:method}, and results presented in section \ref{sec:results} with emphasis on a few case studies. Discussion and conclusions are given in sections  \ref{sec:discuss} and \ref{sec:conclusion}.

\section{Instruments and methods}
\label{sec:method}

\subsection{Instrumentation}
\label{sec:instrument}
The instruments used in this study are:
\begin{itemize}
  \item The Atmospheric Imaging Assembly on board the Solar Dynamics Observatory (AIA/SDO, \citet{lemen14}) provides high spatial resolution (0.6 arcsec/pixel) full disk images with a cadence of 12 s in 10 wavelength bands, including seven EUV, two UV and one visible filter \citep{panesar14}. For this study, images in the 171\r{A} channel are mostly used. This waveband peaks in temperature at $\sim$ 0.8MK and generally gives the best signal above the limb \citep{pagano14}. Images in higher temperature channels are used for further study.
  \item The Large Angle Spectrometric Coronagraph (LASCO) instruments C2/C3 on board the Solar and Heliospheric Observatory (SOHO) \citep{brueckner95} are white light coronagraphs that record the Thompson-scattered emissions from the CME plasma. The spatial resolution of C2 is 11.4 arcsec/pixel, with an useful FOV of 2.2-6.0 \Rs. The spatial resolution of C3 is 56 arcsec/pixel, with an outer FOV limit of 32 \Rs. In 2013/2014 both C2 and C3 usually have a cadence of $\sim$12 minutes.
   \item The Extreme Ultra-Violet Imagers (EUVI) of the Sun-Earth Connection Coronal and Heliospheric Investigation (SECCHI) on board of the paired Solar Terrestrial Relations Observatory (STEREO) Ahead (A) and Behind (B) spacecraft \citep{howard08} gives this study an additional perspective for some of the erupting filaments. As with AIA we use mostly the 171\r{A} channel with the maximum available cadence, usually $\sim$ 5 minutes. During the period of this study, the two STEREO spacecraft were separated by $83^{o}$ on 1st May 2013, and $35^{o}$ on 30th June 2014, aligned on the far side of the Sun from Earth.
  \item The COR2 instruments are twin coronagraphs, aboard STEREO A \& B. They provide a useful field of view of $\sim$3-14\Rs.
\end{itemize}

\subsection{Image Processing}
\label{sec:loop}
Image processing is important to this study as it can reveal structure hidden within the unprocessed data. In particular, we wish to reveal the possible existence of CME front edges in the low corona before, and during, a filament eruption. Two main methods are employed in this work. For the EUV images we use Multi-scale Gaussian Normalization (MGN) processing technique \citep{mgn}. For the white-light coronagraph images we use a Dynamic Separation Technique (DST) to separate the dynamic and quiescent components of the images \citep{morgan10,auto1,morgan2015}.

\subsubsection{MGN Processing}
\label{sec:euv}

\begin{figure*}[!t]
\centering
\includegraphics[width= \figsiz cm]{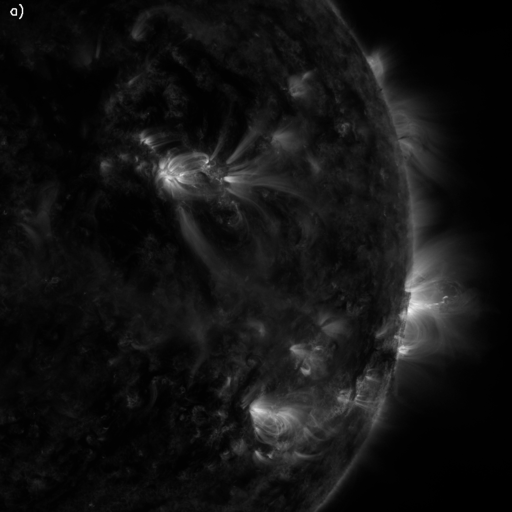}
\includegraphics[width= \figsiz cm]{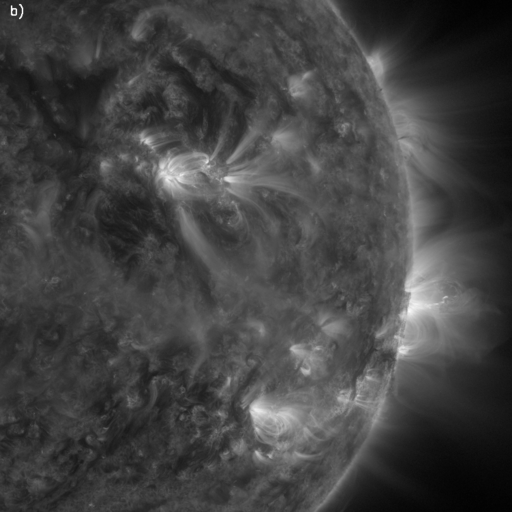}\\
\includegraphics[width= \figsiz cm]{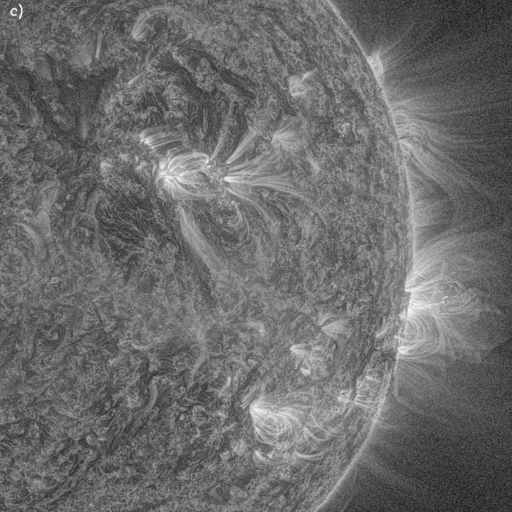}
\includegraphics[width= \figsiz cm]{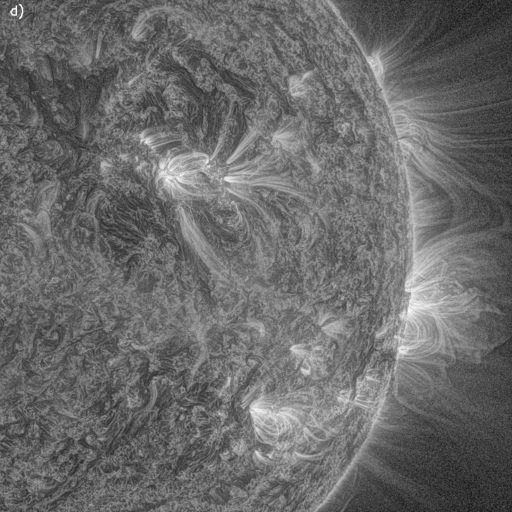}
\caption{An AIA 171\r{A} observation of 2013-08-27 15:54 UT showing the west disk and off-limb region with a) no processing, b) a gamma transform (square root), c) MGN processing, and d) MGN processing on an image combined from five consecutive observations.}
\label{fig:1}
\end{figure*}

Figure \ref{fig:1} shows an example EUV image to demonstrate the application of MGN. EUV images of the corona contain information over a large range of spatial scales and brightness regimes, and an example image is shown in figure \ref{fig:1}a. A common approach to overcome the large contrast between bright and dark regions  is to display a gamma transform (e.g. take the square root of the original image) to reveal more of the low-brightness regions, as shown in figure \ref{fig:1}b. The MGN method normalizes the image at many different spatial scales. This is done by creating a series of new images by applying Gaussian kernels at varying FWHM to the raw AIA image. The mean of these images is taken, and the result is summed together with the original, unprocessed AIA image to create the final result. The method is similar in concept to the Adaptive Histogram Equalisation, but at multiple spatial scales. Information at the finest scales is revealed, while enough of the larger scale information is maintained to provide context. Most important for this study, off-limb structure is enhanced, as shown in figure \ref{fig:1}c. For a full description of the procedure see \citet{mgn}. In this work, in order to further reveal very faint static structure in off-limb regions, a further improvement is made to the method for AIA data by combining exposures. Using the maximum 12 second cadence of AIA, 5 consecutive observations are combined, and the summed images are normalized by the total exposure time of the 5 frames. AIA usually uses a 2 sec exposure. This creates a series of images each combined over an exposure time of 10 seconds. Such a combined image is created for every 1 minute's worth of observation before processing with the MGN procedure, as shown in figure \ref{fig:1}d. At the risk of introducing motion blur for very rapid changes, the structural detail is greatly enhanced, particularly above the limb. Combining of exposures is not used on EUVI images as the cadence is generally too low. As described  in \citet{mgn}; the globally normalized images are processed using a global weighting parameter \textit{h=0.9}. 

\FPset\figsiz{5.0}%figure size for figures 4, 5, 6

\subsubsection{Dynamic Separation Technique (DST) \& coronagraph calibration}
\label{sec:cme}
\begin{figure}[!t]
\centering
\includegraphics[width=6cm]{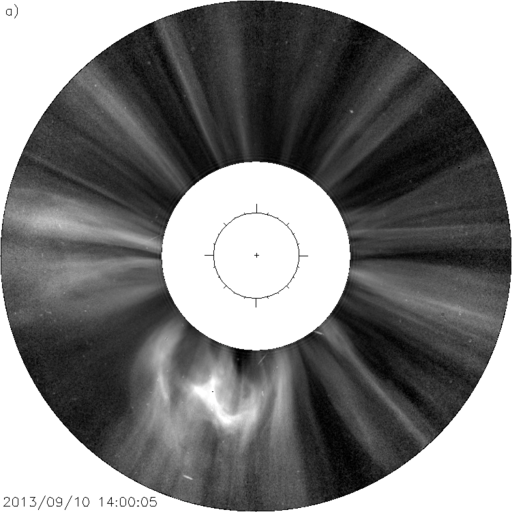}
\includegraphics[width=6cm]{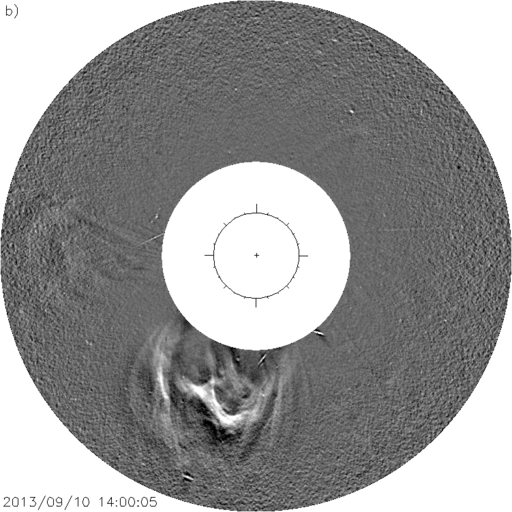}
\caption{CME observed by LASCO C2 on 2013-09-10 14:00. a) C2 image prior to DST, b) separated dynamic component showing the main bright CME, other fainter dynamic events, and noise.}
\label{fig:2}
\end{figure}

In coronagraph images CMEs are not viewed in isolation, but in the presence of the fine structural detail of the quiescent corona such as streamers or coronal holes. In order to accurately track the expansion of CMEs, a processing technique based on spatial and time deconvolution is used that is able to separate the dynamic CME signal from the background quiescent coronal structures \citep{morgan10,auto1}. When applied to observations by LASCO C2 and C3, or SECCHI/COR2, the clear structures of CMEs are revealed despite the presence of background structure that may be several time brighter than the CME. Figure \ref{fig:2} shows an example of the DST applied to a CME observed by LASCO C2 on 2013-09-10. Figure \ref{fig:2}a shows the image prior to separation, and figure \ref{fig:2}b shows the dynamic component of the image post-separation.

For visual inspection of coronagraph images, or a study of kinematics or structure, DST is applied with specific convolving kernel widths in the time and radial dimensions. These are set narrow enough (particularly the spatial, or radial, dimension) to reveal the higher-frequency structures. All the DST images in this work are processed using this bias towards higher frequency detail (or finer spatial detail). For estimates of CME mass, it is necessary to increase the kernel width in the radial direction. This allows lower-frequency components (or broader spatial detail) into the dynamic images, and an improved estimate of mass. See \citet{morgan2015} for a description of this implementation of DST.

All quantitative use of coronagraph data in this work (in particular the mass estimates) have been gained using new calibration techniques described in detail in \citet{morgan2015}. These techniques enable the conversion of the LASCO C2 total brightness measurements (i.e. the most frequent observations made by C2) into k-coronal brightness (\Bk). This is achieved by using the less frequent C2 polarized brightness (\pB) observations as `tie-points', enabling a very robust estimate of the F-corona and stray light backgrounds for subtraction from the total brightness observations. New calibration factors for the \pB\ observations are used (calculated using stars by \citet{morgan2015}), and the radiometric calibration of \citet{colaninno2015} are used (also calculated using stars). 

In the calculation of CME mass from LASCO C2 observations, there are two main sources of uncertainty. One is the radiometric calibration uncertainty, estimated at around 11\%\ by \citet{morgan2015}. Another more dominant error is the lack of information on the distribution of mass along the line of sight. This is a difficult error to quantify. In this work, most of the CMEs are propagating close to the plane of sky, which minimises the line-of-sight error.

\subsection{Bootstrapping}
\label{sec:boot}
Using a point-and-click procedure, the heights of the leading edges of features seen in the processed images of the filament eruption can be tracked through the AIA FOV. The leading edges of the corresponding CMEs as observed by the LASCO C2 and C3 coronagraphs are also recorded using this procedure. These results are combined to form a complete height/time profile of the leading fronts from the starting height above 1\Rs\ to 25\Rs. To use these measurements in the flux-rope model, a fitting of the points to an analytical function is required. This enables the creation of the model at arbitrary times for comparison between the three coronagraphs. In general, the kinematics of events are not well-modelled using a linear height-time fit, especially since our recorded measurements begin in the AIA field of view. To allow for a complicated height-time profile of the CME from eruption to the extended corona, acceleration (\textit{a}) is modeled as a second-order function of time:
\begin{equation}
a(t) = \alpha t^{2} + \beta t + \gamma
\label{eqnkin1}
\end{equation}

A second order function was chosen so that the returned bootstrapping is able to fit the kinematic profile of any CME as accurately as possible. A first order function would not be sufficient as such a function is limited to only allowing for a linear change in acceleration with time; and previous studies have shown that kinematics of different CMEs can take many profiles, triggered by any number of different physical mechanisms \citep{byrne13}.

By integrating, height (\textit{\textbf{r}}) is given by:
\begin{equation}
\textbf{r}(t) = \frac{\alpha}{12} t^{4} + \frac{\beta}{6} t^{3} + \frac{\gamma}{2} t^{2} + v_{0}t + r_{0},
\label{eqnkin2}
\end{equation}
where \textit{\textbf{r{\tiny{0}}}} is the initial height of the loops and $\alpha$, $\beta$ and $\gamma$ are fitted parameters. \textit{x{\tiny{0}}} is set to the initial height of the observed structure prior to the eruption. In order to find estimators for $\alpha$, $\beta$, $\gamma$ and $v_0$ a resampling method called bootstrapping is used. This technique was first introduced by \citet{efron79}, and more recently described in the context of coronal data by  \citet{byrne13}. The implementation of the iterative residual resampling bootstrapping scheme as described in \citet{byrne13}, is as follows:
\begin{enumerate}
  \item An initial fit to the data \textit{r} is obtained using the standard IDL procedure \texttt{mpfitfun.pro}, which performs a Levenberg-Marquardt least-squares fit. This yields the initial model fit \textit{\^{r}} with parameters \textit{\textbf{p}}.
  \item The residuals of the fit are calculated as \textit{e=r-\^{r}}
  \item The residuals are randomly resampled with replacement to give \textit{e*}
  \item The model is then fit to the new data vector \textit{r*=\^{r}+e*} and the parameters \textit{\textbf{p*}} stored. 
  \item Steps 3-4 are repeated many times (e.g., 10,000).
  \item Confidence intervals on the parameters are determined from the resulting distributions.
\end{enumerate}

The bootstrapping scheme is applied to the height-time measurements gained from the manual identification of an eruption's front edge, and the resulting parameters used in equation \ref{eq:2} to yield an analytic function describing the propagation of each event. 

\subsection{Synthetic FR-CME}
\label{sec:frcme}
A synthetic wire-frame model is used to reproduce the large-scale structure of flux-rope-like CMEs. Similar to \citet{thernisien09}, the model consists of a long tubular shape of varying radius, with two foot points anchored to the photosphere which create the body of the CME. This model will be used, along with the kinematic measurements described above, to look in detail at several eruptions. 

The profile of the inner axis (or inner boundary) of the FR model is given initially in a cartesian space by 
\begin{eqnarray}
\label{eq:inneraxis1}
y=\sin(t)[1-\exp(-|t|/\alpha)]\\
z=0.5-0.5\cos(t)\\
x=0,
\label{eq:inneraxis3}
\end{eqnarray}
with parameter $t$ ranging from $-\pi$ to $\pi$. $\alpha$ is a parameter that controls the shape of the inner axis. A helix is formed by
\begin{eqnarray}
x_h=w_h\sin\left(\frac{n_s 2 \pi s}{s_{max}}\right)\\
z_h=w_h\left[ 1+\cos\left(\frac{n_s 2 \pi s}{s_{max}} \right) \right].
\end{eqnarray}
$s$ is the path length along the inner axis, of total length $s_{max}$, with $w_h$ dictating the width of the helix. The number of helix turns is given by $n_s$. A rotation is applied to $x_h$ and $z_h$ so that the helix turns are aligned correctly with the inner axis (rotational plane perpendicular to the CME axis), then the helix coordinates are summed with the inner edge axis as defined by equations \ref{eq:inneraxis1}-\ref{eq:inneraxis3}. The resulting CME shape is centered on the $z$-axis (or the Sun's north pole). Appropriate rotations and scaling are applied to the model CME coordinates, easily giving self-similar radial expansion and views of the CME from the positions of different spacecraft. 

\begin{figure}[!t]
\centering
\includegraphics[width=9cm]{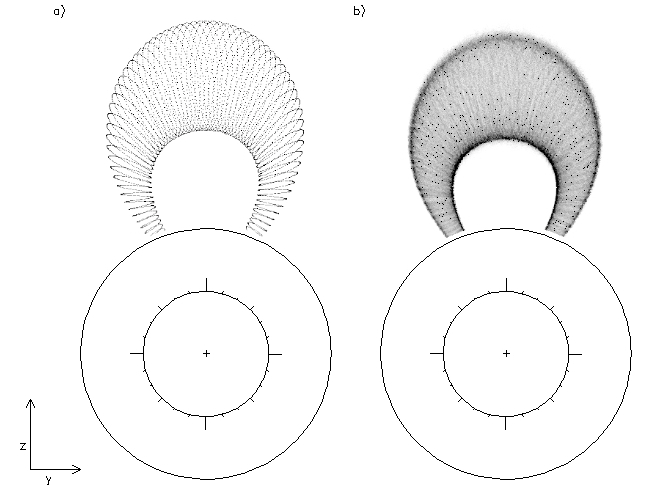}
\caption{Example synthetic flux-rope model. a) A simple rendering of the wire frame structure. b) A volume within a small distance from the surface of the model now contains a random distribution of points to imitate Thompson scattering. Examples rendered as if observed by LASCO C2. Inner and outer circles represent the sizes and positions of the Sun and the C2 occulting disk respectively.}
\label{fig:3}
\end{figure}

By specifying a low number of turns to the helix, and a small number of points, a wire-frame representation of the CME can be given, as shown in figure \ref{fig:3}a. The wire-frame CME shown in figure \ref{fig:3} has been rendered as if observed by SOHO/LASCO C2, with the sizes and locations of the Sun and coronagraph occulter represented by the inner and outer circles respectively. A line-of-sight integrated white light observation of the model CME is formed by increasing the number of points to millions, specifying a large number of rotations in the helix, and also including a small amount of random scatter in the position of points in the model. These points are then treated as individual electrons. For each pixel in the output image, the `electrons' of the model CME which lie within the backprojected solid angle of that pixel's line of sight are summed after weighting with appropriate geometrical factors (as given by \citet{quemerais2002} for example). An example of such an image is given in figure \ref{fig:3}b. This implementation of a FR-CME model is computationally very efficient, and many millions of points can be used within the CME model. Pixels in the output image thus contain contribution from many thousands of points which represent individual scattering electrons in the model CME.

Applying the height-time function (equation \ref{eqnkin2}) to the expansion of the model CME recreates the entire eruption as well as extrapolating the heights at time intervals smaller than that of the coronagraph cadence, filling in the time gaps between coronagraph images in the sequence. Other input parameters for the model are the central longitude and latitude of the CME origin, which are assumed to be those of the filaments as seen by AIA/EUVI. The orientation and width of the synthetic flux-rope is adjusted through manual comparison of the synthetic and real observations.

\section{Results}
\label{sec:results}

\begin{figure*}[!t]
\centering
\includegraphics[height=\figsiz cm]{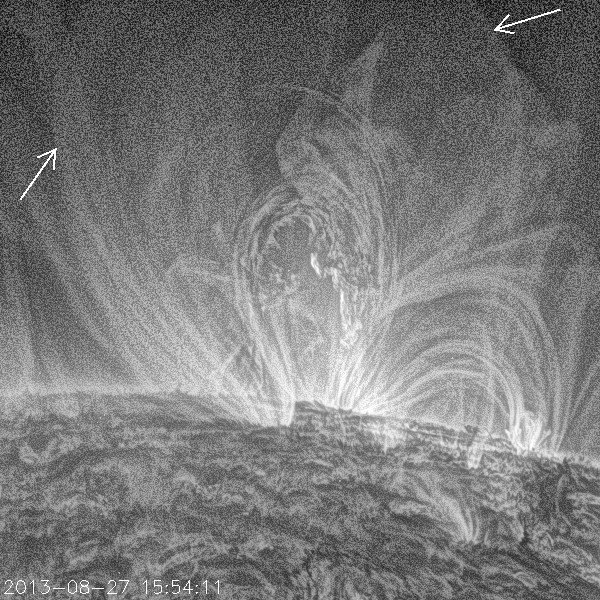}
\includegraphics[height=\figsiz cm]{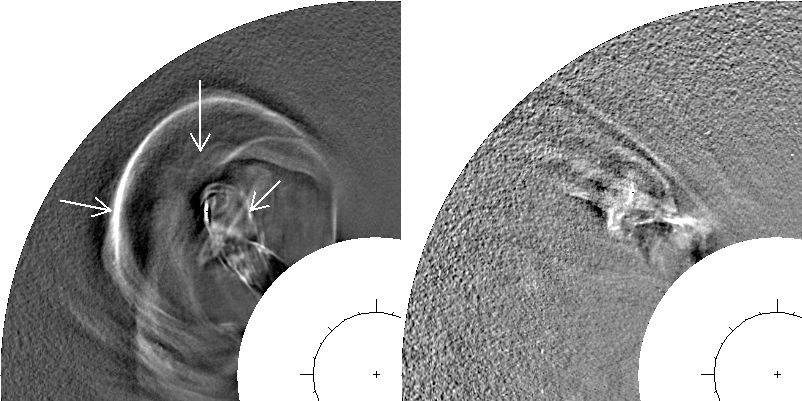}
\caption{(a) AIA 171\AA\ image of a filament during eruption. Arrows point to a system of large loops overlying the filament, which are in the process of erupting along with the underlying filament. (b) A classic 3-part structure CME with arrows pointing to the front edge, cavity and inner core. (c) A CME without clear 3-part structure.}
\label{fig:exampleloops}
\end{figure*}

During a 14-month period from 2013/05/03 to 2014/06/30 a total of 221 filament eruptions were identified which were centered near the limb in AIA images. For each event, a time-sequence of MGN-processed images are created using AIA and STEREO A or B (as appropriate) EUVI  observations. The appropriate time sequence of LASCO C2, C3 and STEREO A or B COR2 observations are DST-processed for each event. These sets of processed images are examined for specific observational characteristics: the existence of a set of loop structures lying above the pre-eruptive filament in EUV images which subsequently erupt with the filament, and a clear 3-part CME structure in the coronagraph observations. Figure \ref{fig:exampleloops}a shows an example of a pre-existing system of loops above a filament which is in the process of erupting with the underlying filaments. Figure \ref{fig:exampleloops}b shows an example of a CME which has a clear 3-part structure, and figure \ref{fig:exampleloops}c is an example of a CME which fails to show a clear 3-part structure.

\FPset\figsiz{5.0}%figure size for figures 4, 5, 6

Of the 221 events, 88 eruptions failed to reach the coronagraph field of view. These 88 events will be discarded for the remainder of this study. Each of the remaining 133 events are categorized according to the following criteria:
\begin{itemize}
\item Category 1: Filament eruptions that are seen to have pre-existing system of loops, observed in EUV, overlaying them prior to and during eruption, which are followed by a clear 3-part CME, with the leading front clearly arising from the pre-existing loops. In this category there are 31 events (23.3\% of the total 133).
\item Category 2: Filament eruptions that did not show pre-existing loop structures, but which did have a clear 3-part CME. In this category there are 61 events (45.9\%).
\item Category 3: Filament eruptions that did show pre-existing loop structures, but no clear 3-part CME. In this category there are 2 events (1.5\%).
\item Category 4: Filament eruptions with neither pre-existing loop structures nor a clear 3-part CME. In this category there are 39 events (29.3\%).
\end{itemize}
Table 3 in the Appendix gives details of all 221 events, along with their category. The following sections discuss each category and show details of case studies relating to each category. 

\subsection{Category 1}
\label{sec:cat1}
\begin{figure*}[!t]
\centering
\includegraphics[width=\figsiz cm]{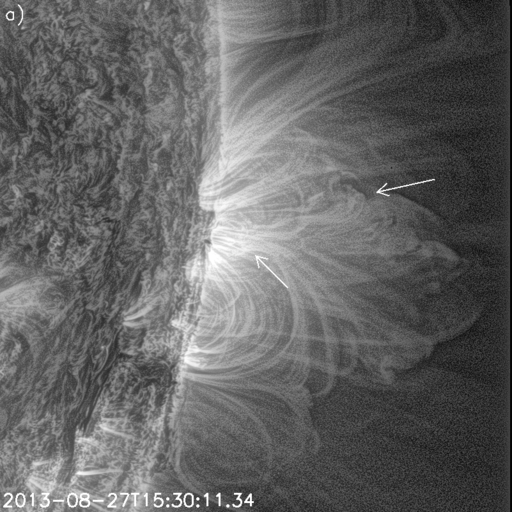}
\includegraphics[width=\figsiz cm]{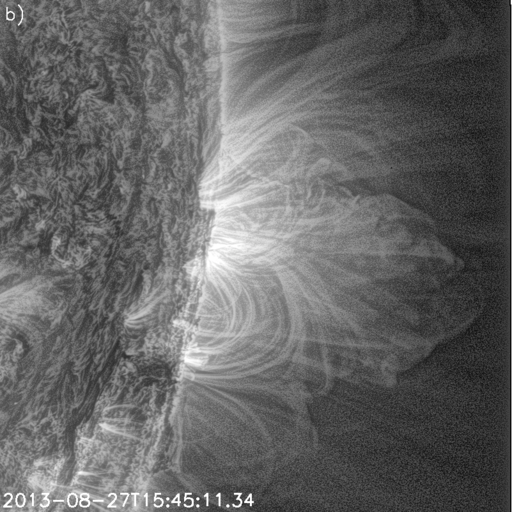}
\includegraphics[width=\figsiz cm]{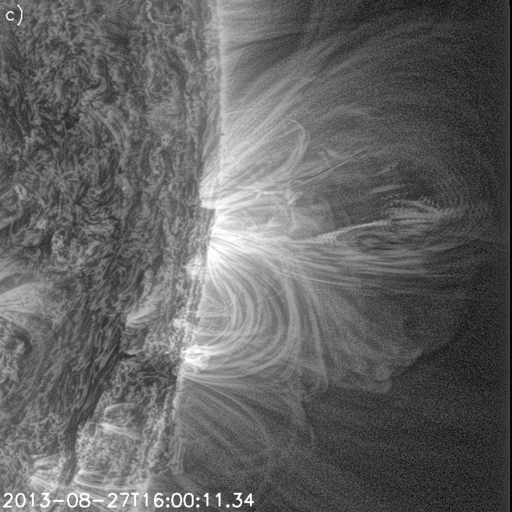}\\
\includegraphics[width=\figsiz cm]{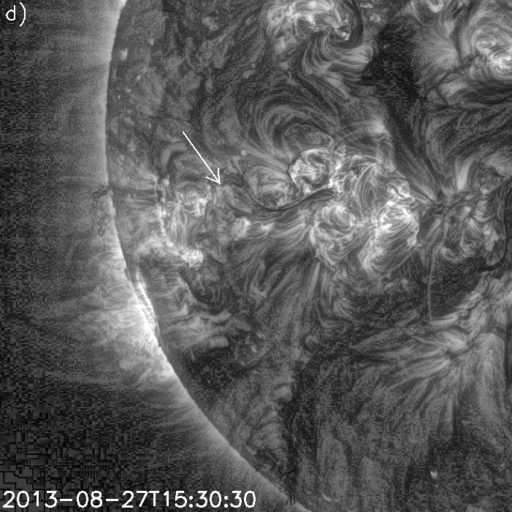}
\includegraphics[width=\figsiz cm]{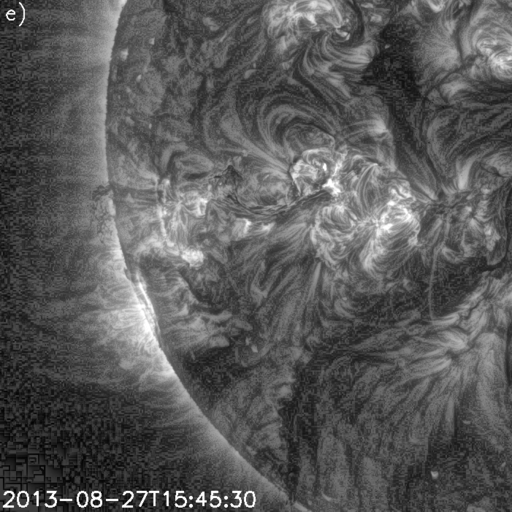}
\includegraphics[width=\figsiz cm]{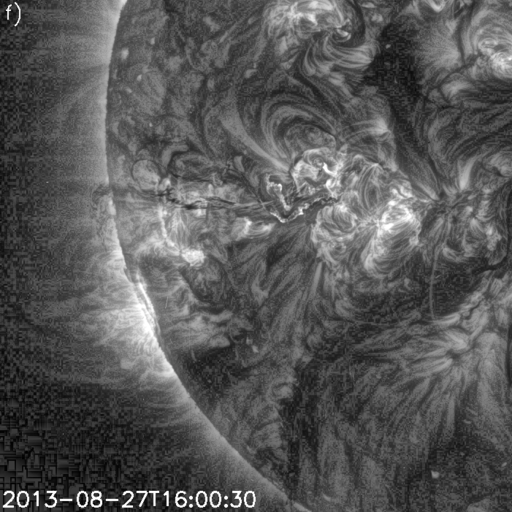}\\
\caption{MGN processed EUV images of the 2013-08-27 event in 15 minute intervals from 15:30-16:00 UT, showing the looped structures above the filaments. Top row (a-c) AIA images, bottom row (d-f) the corresponding EUVI A images. Animations of this event are available online. See Appendix.}
\label{fig:4}
\end{figure*}

\begin{figure*}[!t]
\centering
\includegraphics[width=\figsiz cm]{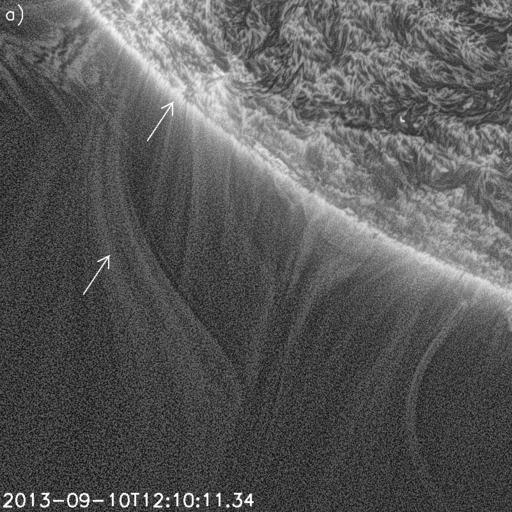}
\includegraphics[width=\figsiz cm]{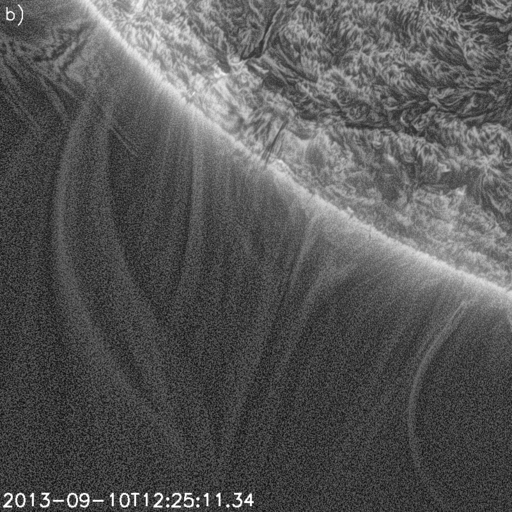}
\includegraphics[width=\figsiz cm]{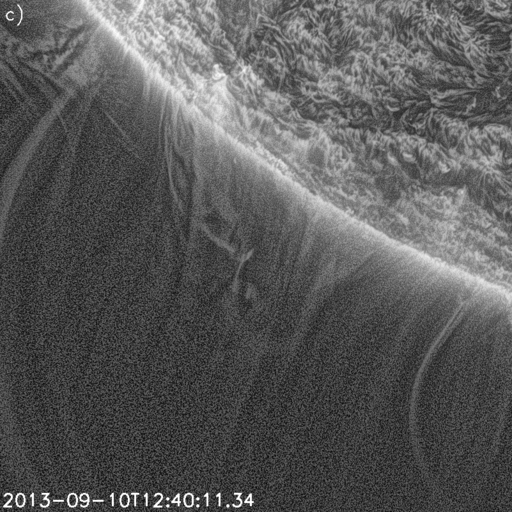}\\
\includegraphics[width=\figsiz cm]{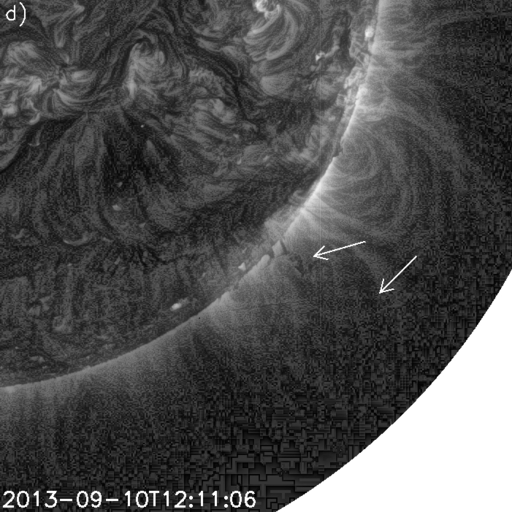}
\includegraphics[width=\figsiz cm]{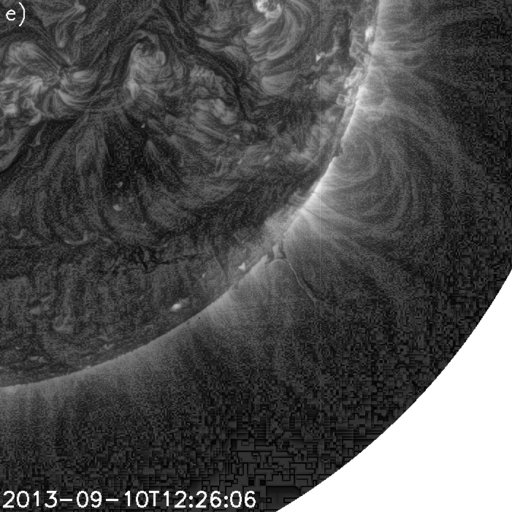}
\includegraphics[width=\figsiz cm]{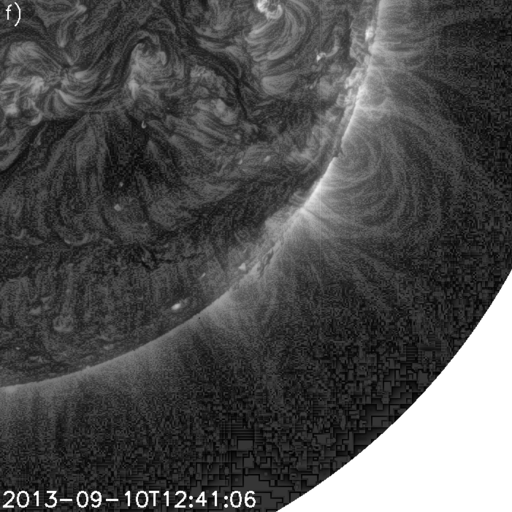}\\
\caption{As figure \ref{fig:4}, but for the 2013-09-10 event in 15 minute intervals from 12:10-12:40 UT, using AIA (top row) and EUVI B (bottom row). Animations of this event are available online. See Appendix.}
\label{fig:5}
\end{figure*}

\begin{figure*}[!t]
\centering
\includegraphics[width=\figsiz cm]{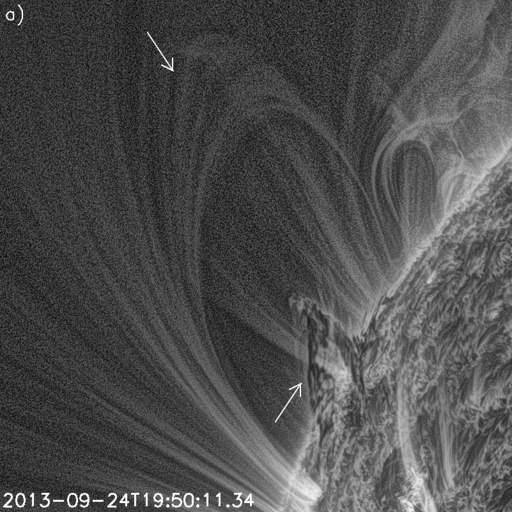}
\includegraphics[width=\figsiz cm]{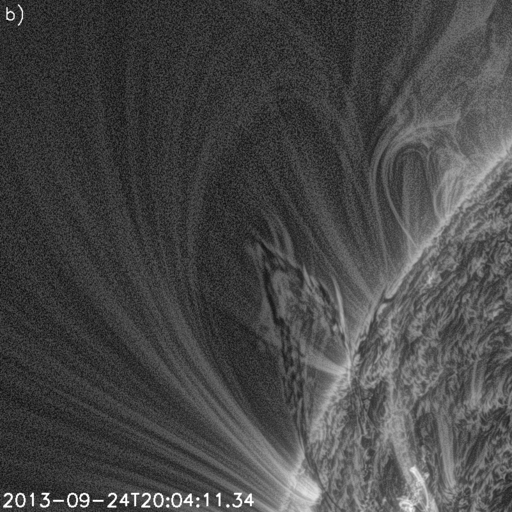}
\includegraphics[width=\figsiz cm]{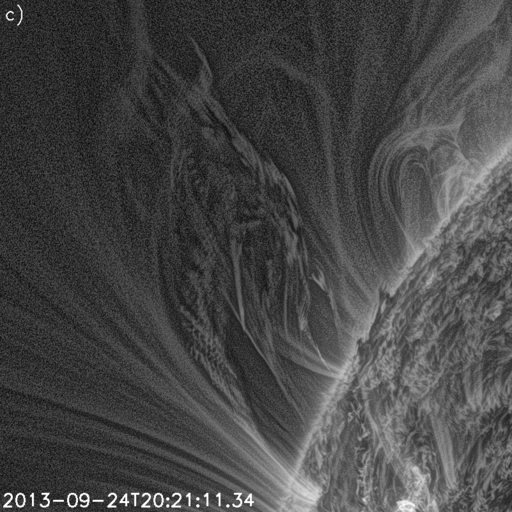}\\
\includegraphics[width=\figsiz cm]{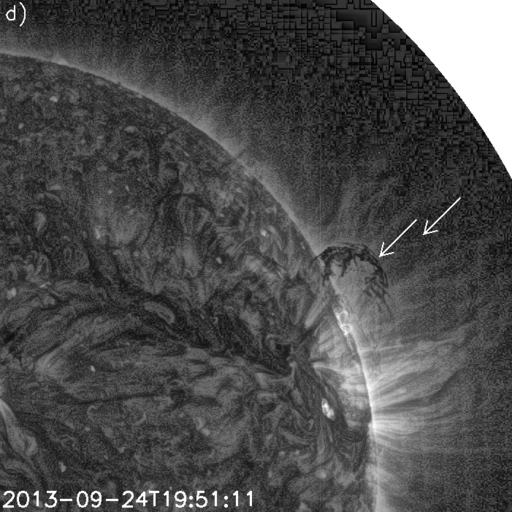}
\includegraphics[width=\figsiz cm]{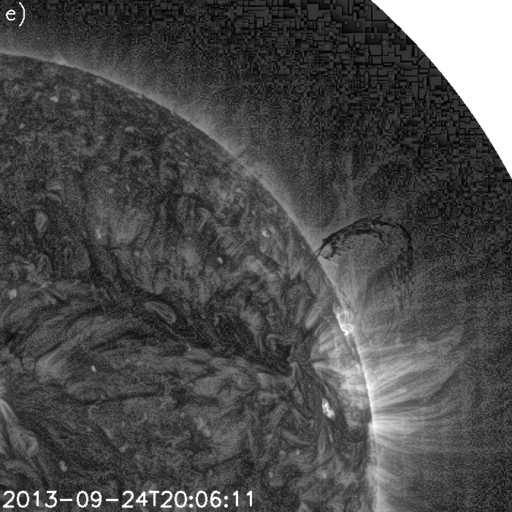}
\includegraphics[width=\figsiz cm]{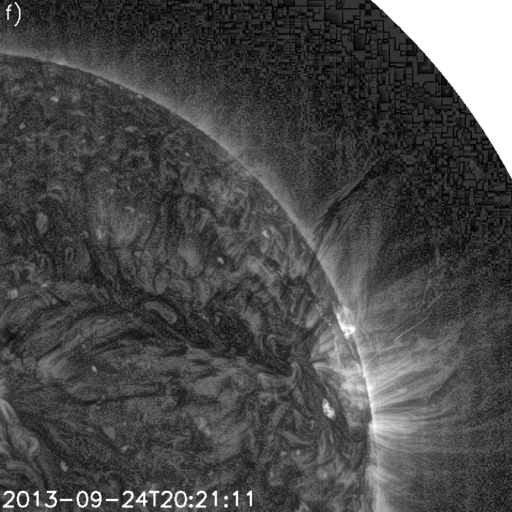}\\
\caption{As figure \ref{fig:4}, but for the 2013-09-24 event in 15 minute intervals from 19:50-20:20 UT, using AIA (top row) and EUVI B (bottom row). Animations of this event are available online. See Appendix.}
\label{fig:6}
\end{figure*}

This section, dedicated to 3 examples of Category 1 events, shows the detailed analysis which shows the careful selection of events as ones which possess a pre-existing system of loops, and, more importantly, the association of these loops with the bright front edge of a 3-part CME. The three selected events are:\\
\noindent
\emph{Event 1, 2013/08/27:} The eruption begins at 15:30UT. An active region filament eruption, emerging from the west limb, is seen near the equator in the AIA FOV. See figure \ref{fig:4}.\\
\emph{Event 2, 2013/09/10:} Begins 12:00UT. Erupting near the south-east limb in the AIA FOV. See figure \ref{fig:5}.\\
\emph{Event 3, 2013/09/24:} Begins 20:00UT. Erupting on the north-east limb in the AIA FOV. See figure \ref{fig:6}.

\FPset\figsiz{5.0}%figure size for figures 4, 5, 6

Although the primary EUV observations are made from SDO/AIA, STEREOs EUVI instruments were also used to confirm the existence of pre-existing loops and to gain a better idea of their three-dimensional (3D) shape. Based on the relative positions of the two STEREO spacecraft at the time, EUVI-A was used for the 2013-08-27 event, and EUVI-B for the other two. Processing reveals loop-like structures existing above the filament prior to the eruption for all three case events, and which erupt and expand in front of the erupting filament.

\begin{figure}[!t]
\centering
\includegraphics[width=\figsiz cm]{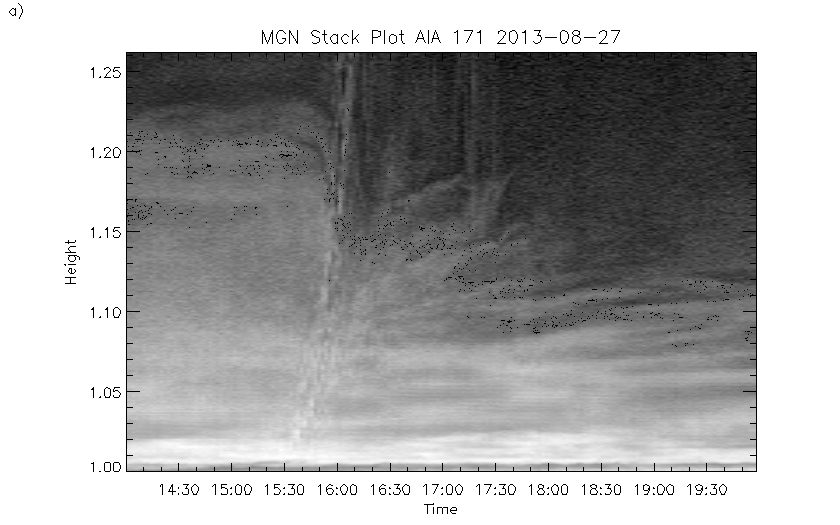}
\includegraphics[width=\figsiz cm]{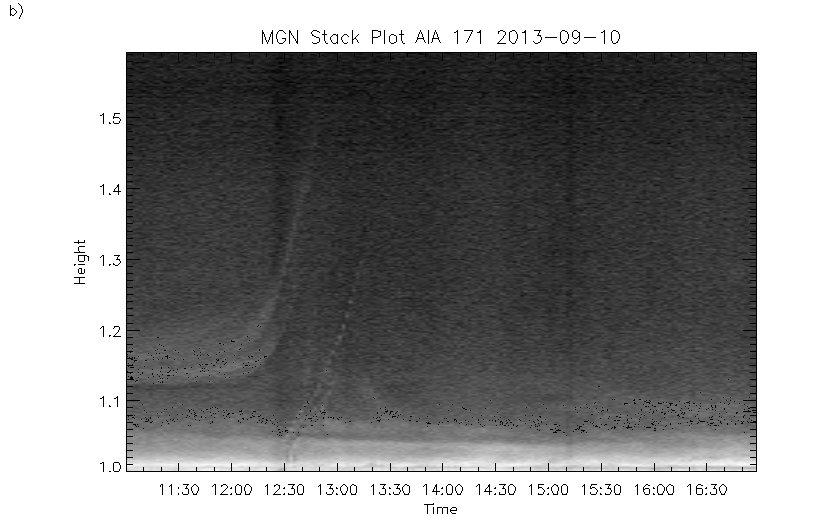}
\includegraphics[width=\figsiz cm]{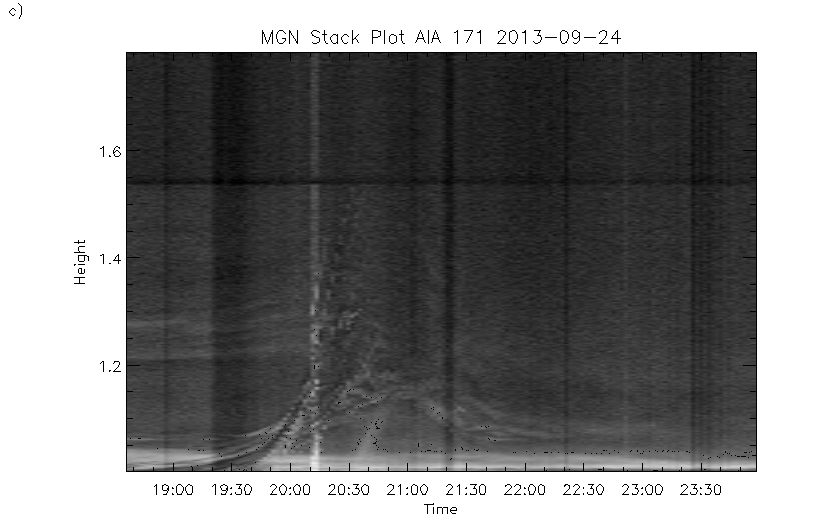}
\caption{Height-time stack plots showing three example eruptions corresponding to Category 1 events. (a): 2013/08/27, (b): 2013/09/10 and (c): 2013/09/24 event.}
\label{fig:7}
\end{figure}

Using the processed images from AIA, height-time stack plots were made for each of the events. These plots were created by selecting a single array of pixels, starting at the edge of the solar disk and tracing a straight line outwards, following the path of each eruption. The stack plots are used to track the height of the leading loops, and the filaments, above the solar surface with time (figures \ref{fig:7}a-c). Both the 2013/09/10 and 2013/09/24 events (figures \ref{fig:7}b and c) show very clearly a bright loop structure existing above the erupting filament for at least several hours prior to eruption, at heights $\sim$0.2\Rs\ above the limb. These consequently erupt with the filament. The 2013/08/27 event (figure \ref{fig:7}a) originates from an equatorial active region, with many other bright structures in the same region which conceals any clear view of the leading loops in the stack plot. However there is a large amount of structure prior to the eruption at heights $\sim$1.2\Rs\  (figure \ref{fig:7}a) that disappear with the eruption. In the movie, the front system of loops is more clearly seen to expand out with the erupting filament. Movie clips for each of these three events are available online. See Appendix for details.

The profile of the eruptions seen in the stack plots (Figure \ref{fig:7}) can give clues to the triggering mechanisms. The stack plot for the 2013/09/10 event shows that the loop structure exists for several hours before the eruption. The loop structure is visible from as early as 08:00 at heights 1.14-1.20\Rs, up to the eruption at 12:30. The loops themselves begin to move at around 12:10, approximately 20 minutes earlier than the filament. This could be evidence of the breakout model triggering magnetic reconnection in the overlaying field holding the filament in equilibrium. This would decrease the field strength of the overlaying loop structures, causing them to expand first. The subsequent decrease in magnetic pressure in the overlaying fields triggers the eruption of the filament. The 2013/09/24 loops also exist long prior to the eruption, visible from at least as early as 17:00. Existing at heights of 1.20-1.26\Rs\, they share a common apparent thickness as the loop structures of the previous event. Unlike the previous event, the filament appears to erupt before the loops begin to expand - possible evidence of reconnection occurring below the filament/cavity flux tube, which is indicative of the tether-cutting model or of a kink-instability.

\FPset\figsiz{7.0}%figure size for figure 7

\begin{figure}[!t]
\centering
\includegraphics[width=\figsiz cm]{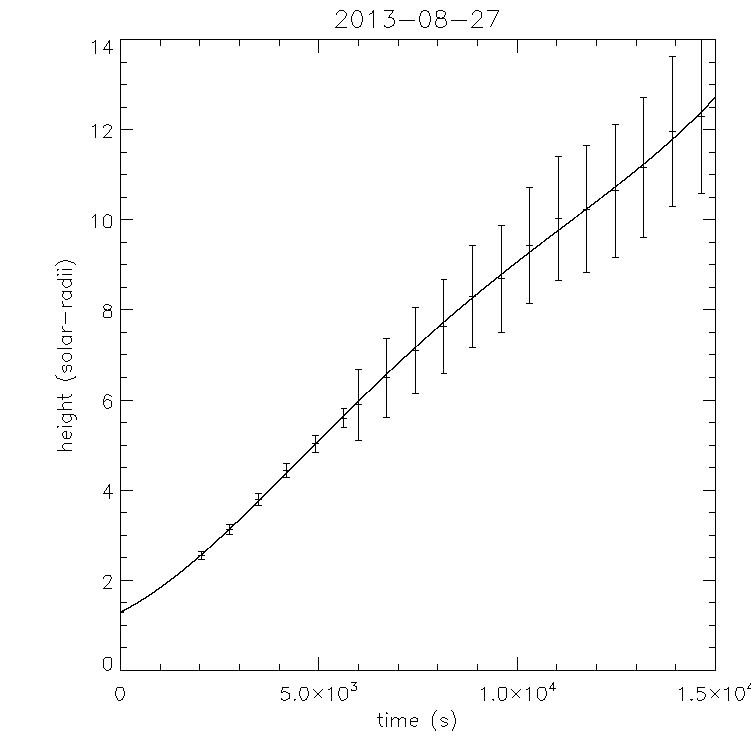}
\includegraphics[width=\figsiz cm]{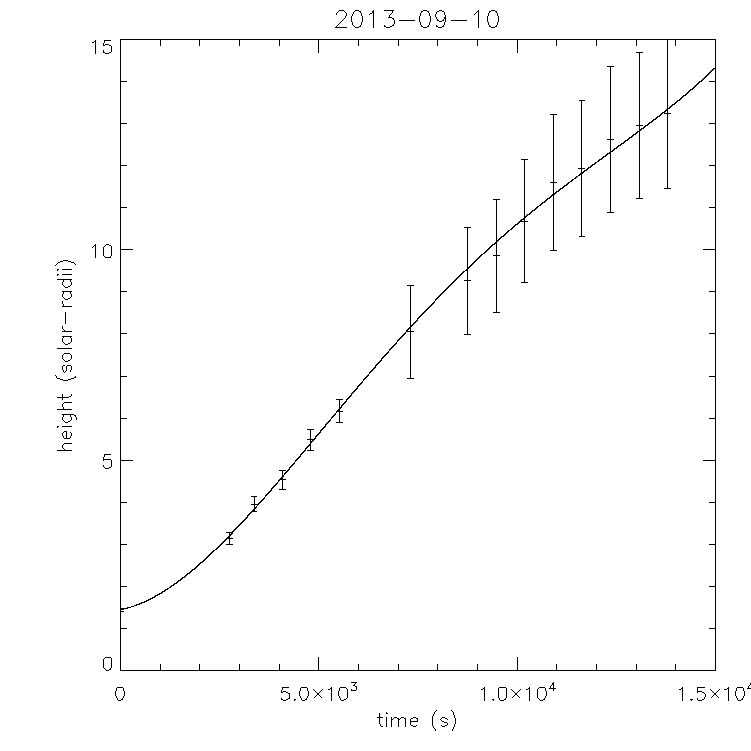}
\includegraphics[width=\figsiz cm]{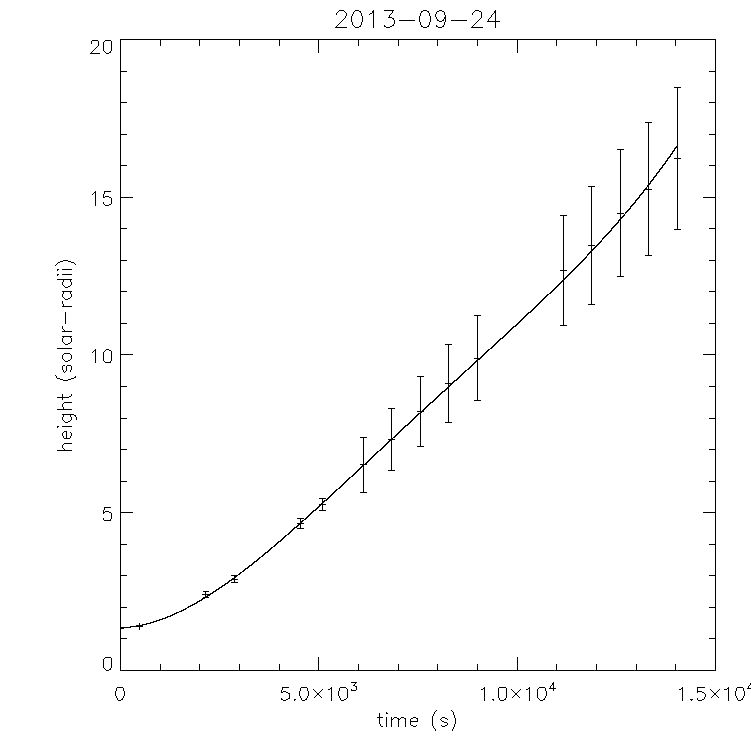}
\caption{Height of the flux rope leading front vs. time from AIA, C2 and C3 data for (a) 2013/08/27, (b) 2013/09/10, and (c) 2013/09/24 events. The fitted kinematics is the solid line.}
\label{fig:8}
\end{figure}

Having shown the existence of pre-existing loop structures which erupt with the filaments, we aim to show that these expanding loops become the bright leading edges of the CMEs as viewed by LASCO C2. For this, we employ a flux-rope density model, with kinematics constrained by both EUV and coronagraph observations. The expansion of the leading fronts of these flux rope loop structures are manually tracked outwards through the AIA FOV, and the same manual tracking records the position of the leading front of the CME through the C2/C3 FOVs. Combining these data gives a total height-time profile of the expanding flux-rope. The bootstrapping method is applied to produce an accurate estimator for the complete kinematic profile of the eruptions. Figures \ref{fig:8}a-c show the height data with the bootstrapping fit, and table 1 contains the information on the starting condition of the loop structures for the three case events (start time of the eruption, latitude and longitude of the filament on the solar surface, and the initial height of the loops above the limb) as well as the bootstrapped fitting parameters ($\alpha$, $\beta$, $\gamma$ and \textit{v{\tiny{0}}}) for each of the case events (see equations \ref{eqnkin1} and \ref{eqnkin2}). 
Given the high number of degrees of freedom, the uncertainties for velocity and acceleration are calculated using reduced chi squared goodness of fit. The mean uncertainties for category 1 events are 7.66 km.$s^{-1}$ for velocity and 5.84 m.$s^{-2}$ for acceleration.

\begin{table}[t]
\centering
\caption{Date, start time, central Carrington position and the initial height (heliocentric) of the pre-existing loop structures prior to eruption. $\alpha$, $\beta$ and $\gamma$ are constants for the polynomial as defined in equations \ref{eqnkin1} and \ref{eqnkin2}, and \textit{v$_0$} is the bootstrapped estimate for the initial velocity of the erupting filament.}
\begin{tabular}{ccccc}%rrrr}
\hline\hline
Date & Time & Lat. & Lon. & Ht. (\Rs)
\\\hline
2013-08-27 & 15:30 & $-10^{o}$ & $101^{o}$ & 1.35 \\
2013-09-10 & 12:30 & $-70^{o}$ & $-35^{o}$ & 1.72 \\
2013-09-24 & 20:00 & $35^{o}$ & $-70^{o}$ & 1.41 \\
\hline\\\hline\hline
$\alpha$ ($10^{-9}$ km.$s^{-4}$) & $\beta$ ($10^{-5}$ km.$s^{-3}$) & $\gamma$ (km.$s^{-2}$) & \textit{v$_0$} (km.$s^{-1}$) & \\
\hline
3.06 & -4.85 & 0.16 & 305.0 & \\
4.83 & -8.39 & 0.3 & 130.0 & \\
5.91 & -8.93 & 0.33 & 23.8 & \\
\hline
\end{tabular}
\end{table}

\FPset\figsiz{5.0}%figure size for figure 7

\begin{figure*}[]
\centering
\includegraphics[scale=\figsiz]{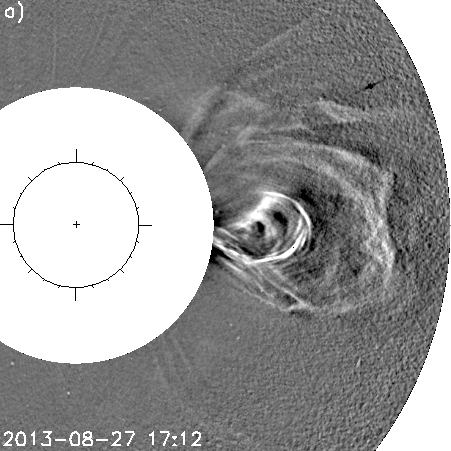}
\includegraphics[scale=\figsiz]{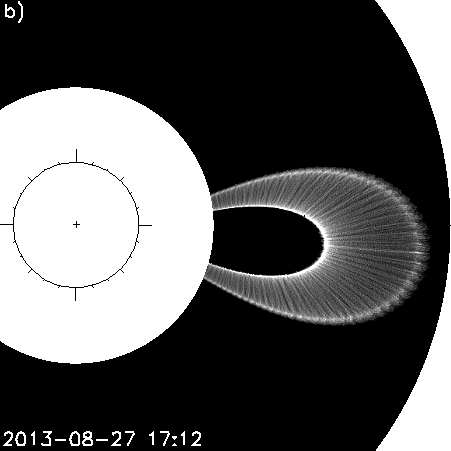}\\
\includegraphics[scale=\figsiz]{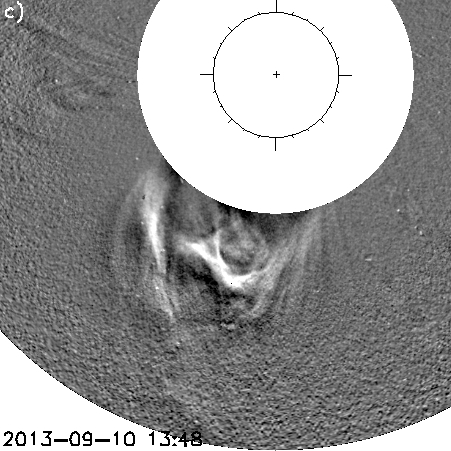}
\includegraphics[scale=\figsiz]{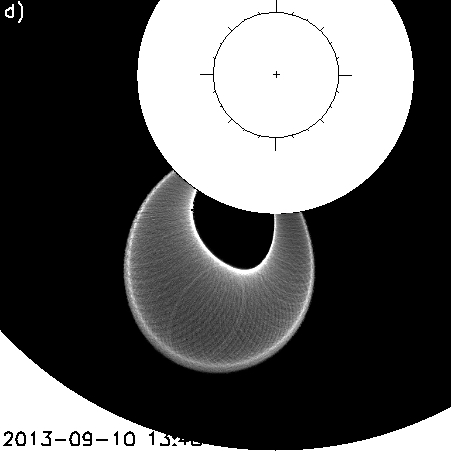}\\
\includegraphics[scale=\figsiz]{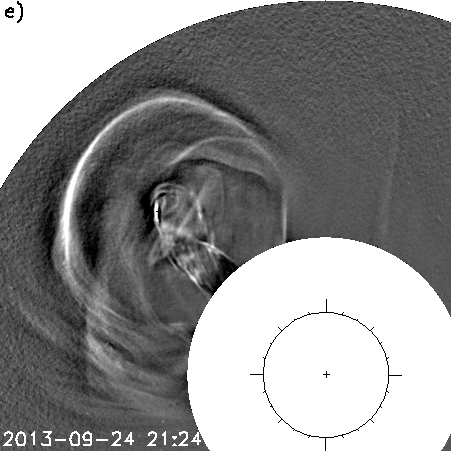}
\includegraphics[scale=\figsiz]{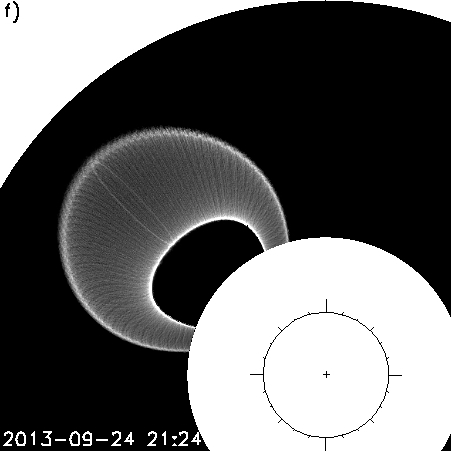}
\caption{Left column: Dynamic separated, white light LASCO C2 coronagraph images of the CMEs corresponding to each case event of category 1. Right column: The synthetic wire-frame flux rope model of that CME. a,b) Event 1: 2013-08-27.  c,d) Event 2: 2013-09-10 e,f) Event 3: 2013-09-24.}
\label{fig:9}
\end{figure*}

\begin{figure*}[]
\centering
\includegraphics[scale=\figsiz]{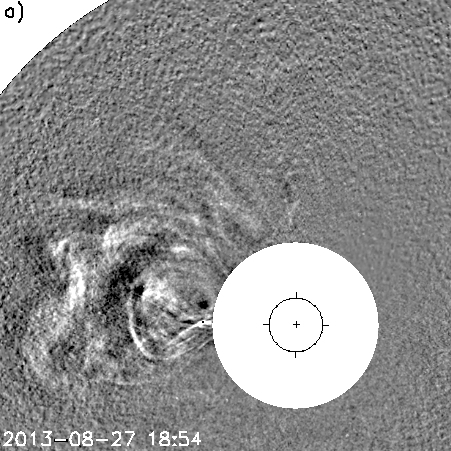}
\includegraphics[scale=\figsiz]{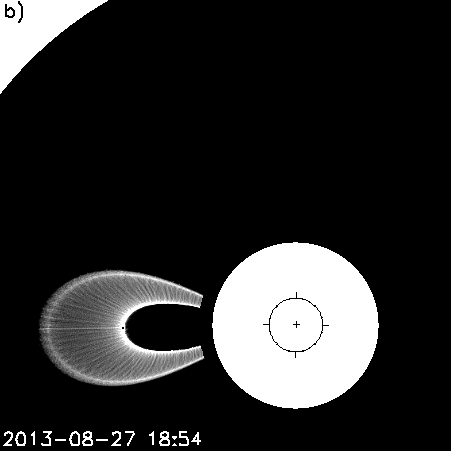}\\
\includegraphics[scale=\figsiz]{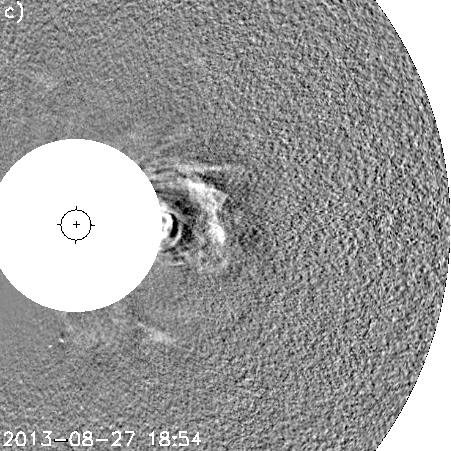}
\includegraphics[scale=\figsiz]{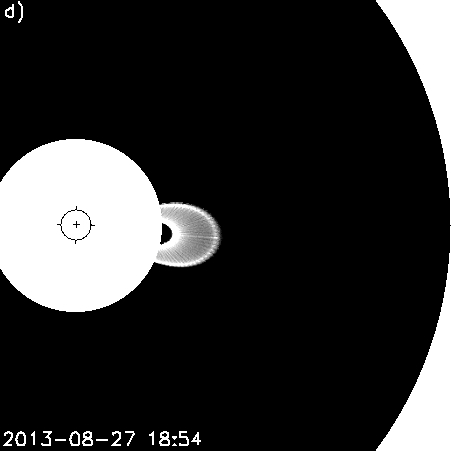}\\
\includegraphics[scale=\figsiz]{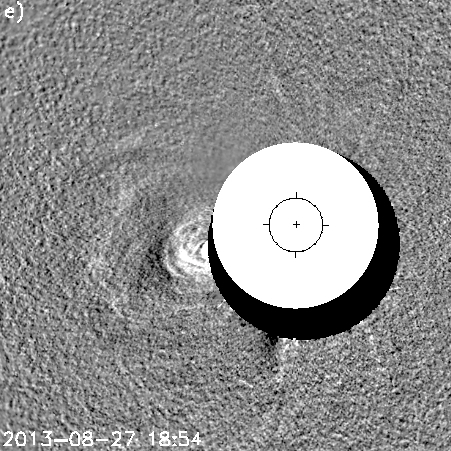}
\includegraphics[scale=\figsiz]{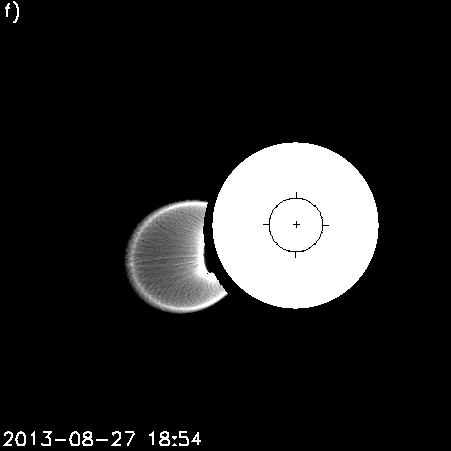}
\caption{Images of the observed 2013-08-27 event (left column) and corresponding model images (right column) for (a,b) STEREO-A COR2. Middle(c,d): LASCO C3. Bottom row(e,f):STEREO-B COR2.}
\label{fig:10}
\end{figure*}

\begin{figure*}[]
\centering
\includegraphics[scale=\figsiz]{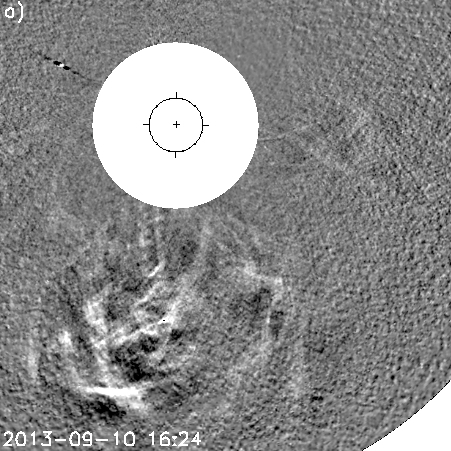}
\includegraphics[scale=\figsiz]{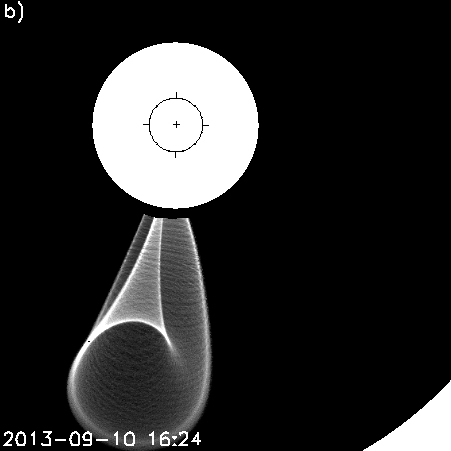}\\
\includegraphics[scale=\figsiz]{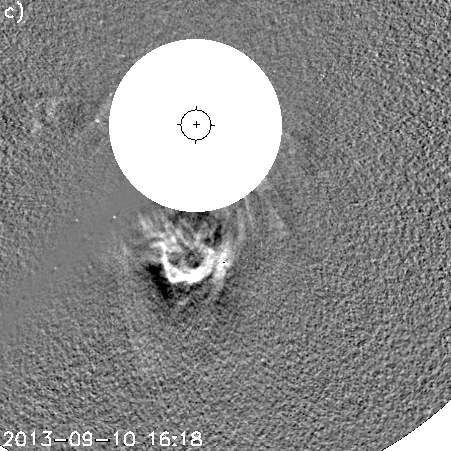}
\includegraphics[scale=\figsiz]{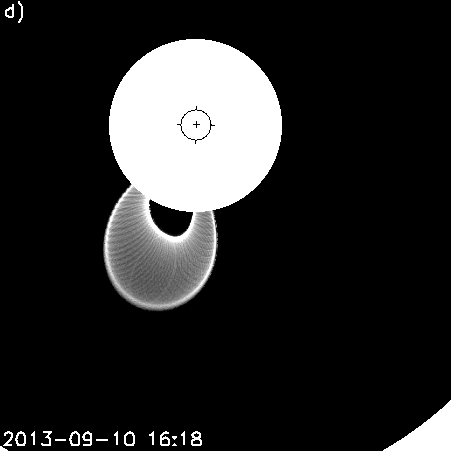}\\
\includegraphics[scale=\figsiz]{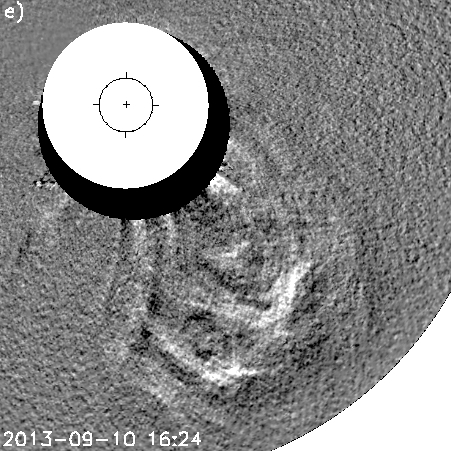}
\includegraphics[scale=\figsiz]{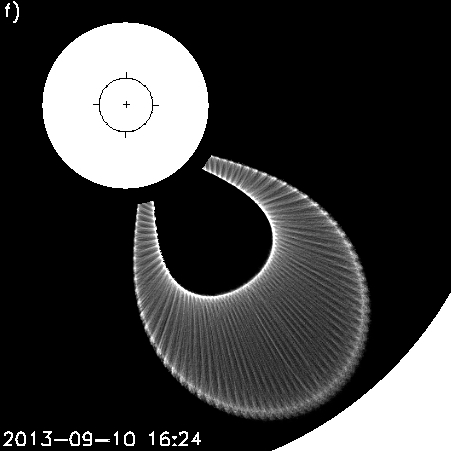}
\caption{Images of the observed 2013-09-10 event (left column) and corresponding model images (right column) for (a,b) STEREO-A COR2. Middle(c,d): LASCO C3. Bottom row(e,f):STEREO-B COR2.}
\label{fig:11}
\end{figure*}

\begin{figure*}[]
\centering
\includegraphics[scale=\figsiz]{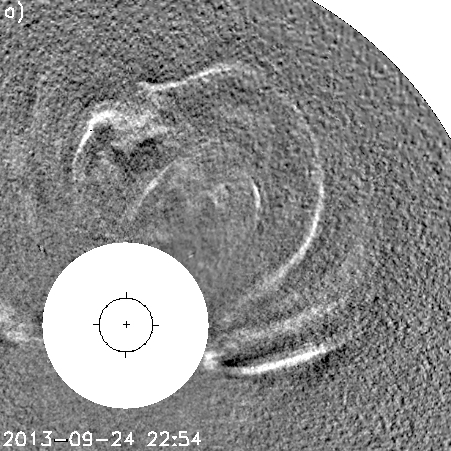}
\includegraphics[scale=\figsiz]{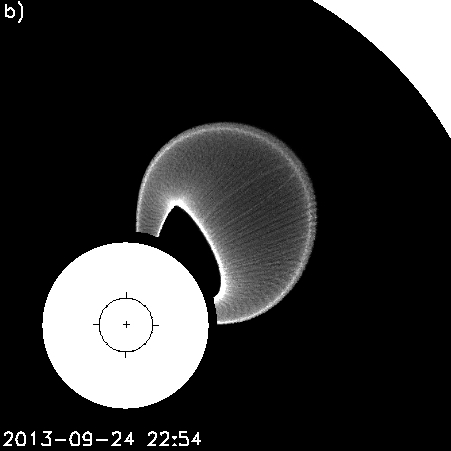}\\
\includegraphics[scale=\figsiz]{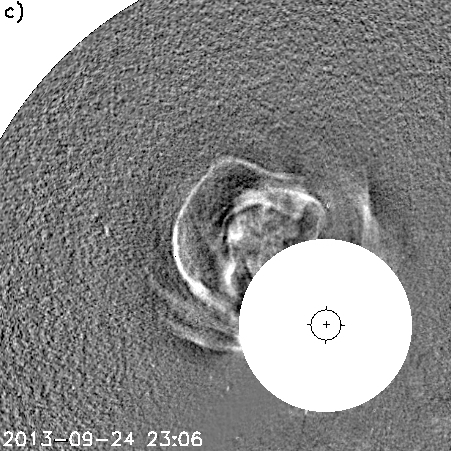}
\includegraphics[scale=\figsiz]{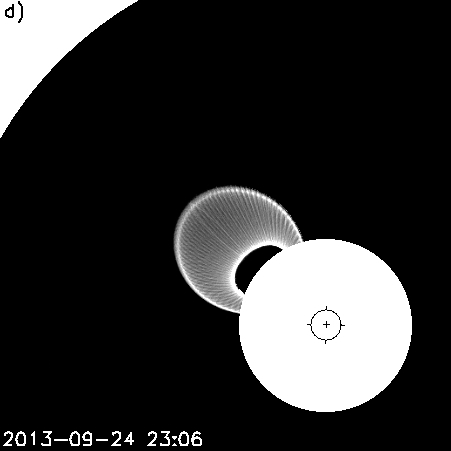}\\
\includegraphics[scale=\figsiz]{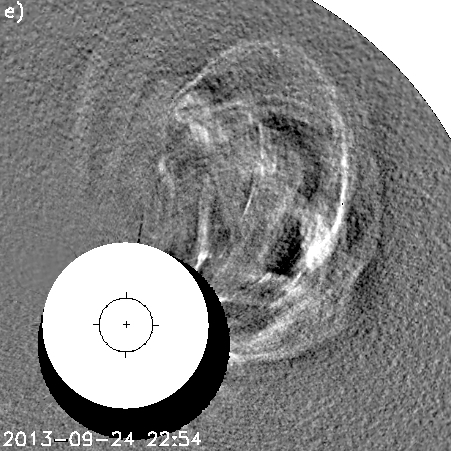}
\includegraphics[scale=\figsiz]{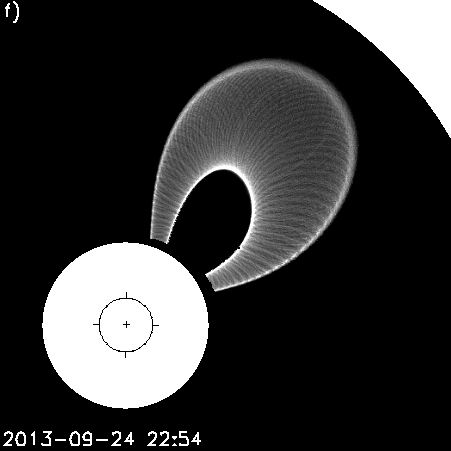}
\caption{Images of the observed 2013-09-24 event (left column) and corresponding model images (right column) for (a,b) STEREO-A COR2. Middle(c,d): LASCO C3. Bottom row(e,f):STEREO-B COR2.}
\label{fig:12}
\end{figure*}

The important parameters of the flux-rope density model are the central longitude, central latitude, peak height, leg-to-leg width relative to height, peak diamater of flux tube and orientation around the central axis. For this analysis work, the central longitude and latitude are fixed by the central longitude and latitude of the erupting filament. The peak height (as a function of time) is given by equation \ref{eqnkin2} and the fitted kinematic parameters listed in table 1. The orientation around the central axis is adjusted to give the best agreement between model and coronagraph observations (from LASCO C2 and COR2A \& B). The leg-to-leg width and peak diameter of the flux tubes are kept fixed, but proportional to the peak height (thus maintaining radial expansion). In this way, we show that the CME front edge is consistent with a flux-rope structure, and that the outer layer of the flux rope structure is consistent with the observed kinematics of the expanding cavity above the erupting filament.  Figures \ref{fig:9}a-f show the C2 images of the coronagraph data (left) and the corresponding synthetic white light images taken from the flux-rope density model (right). Likewise, the left columns of figures \ref{fig:10}, \ref{fig:11} and \ref{fig:12} show the coronagraph images of COR2A, C3 and COR2B (top to bottom) taken at a similar time, and the right columns show the corresponding synthetic CME images.

The 2013/08/27 CME is seen by LASCO to erupt off the west limb from an active region close to the equator. A good visual agreement is seen between the model leading edge and the CME front at the corresponding times for all observers (figure \ref{fig:10}). Given the position of STEREO B, the CME is erupting away from the observatory. The leading edge of the CME is very faint, but still visible in the STEREO B images, and is in good agreement with the model. As shown by highest initial velocity (\textit{v{\tiny{0}}}=305km.$s^{-1}$) as estimated by the bootstrapping scheme, this event is more impulsive than the other two case studies. However, by the time the CME is visible in the C2 FOV, and thereafter, the CME propagates with velocities very similar to that of the 2013/09/10 event. Both CMEs reach a height of 15\Rs\ approximately 4:30hr after erupting. Reviewing the magnetogram data from SDO/HMI in the days leading up to this eruption shows that the regions magnetic field was indeed highly active. However by the start of 2013/08/27 the active region had already moved to the solar limb, so it is not possible to comment on the evolution of the magnetic field inside of the active region soon prior to eruption.

The 2013/09/10 example is probably associated with a breakout trigger. There was an unfortunate large datagap in COR2 B from 13:54UT to 16:42UT. The leading front of the CME becomes faint very quickly, especially by the time it emerges into the C3 FOV. However the bright core of prominence material and the dark cavity are still both clearly visible. A comparison of the synthetic CME and the coronagraph data shows a good visual agreement in figure \ref{fig:11}.

The CME of 2013/09/24 has the brightest, most clear structure throughout the entire eruption in all views (figure \ref{fig:12}). In this case the central core of the CME can easily be identified as the filament eruption as seen by AIA (figure \ref{fig:6}) as much of the structural characteristiccs are maintained in the C2 images. The clear structure of the CME made the leading edge height easy to identify manually, giving confidence in the heights and kinematics: the height-time plot and the bootstrapped kinematics show the best correlation of all events. This event also shows the best visual agreement between CME and synthetic images.

\FPset\figsiz{0.3}%figure size for figure 7

\clearpage
\subsection{Category 2}
\label{sec:cat2}
\begin{figure*}[t]
\centering
\includegraphics[width=\figsiz cm]{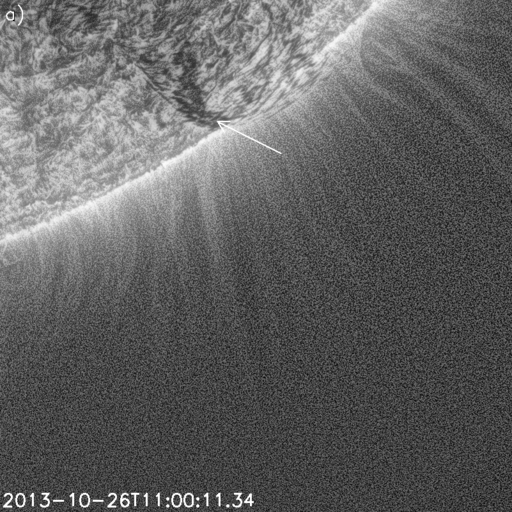}
\includegraphics[width=\figsiz cm]{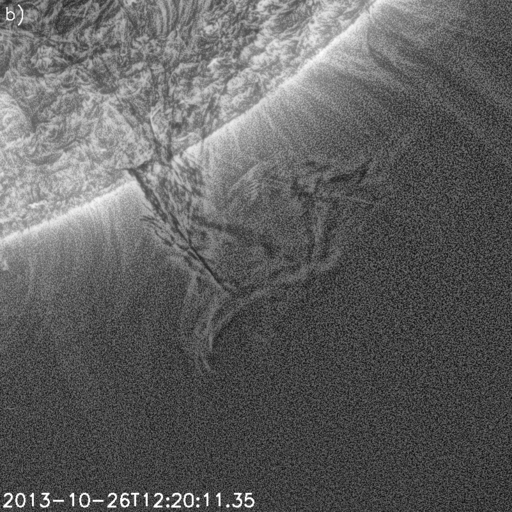}
\includegraphics[width=\figsiz cm]{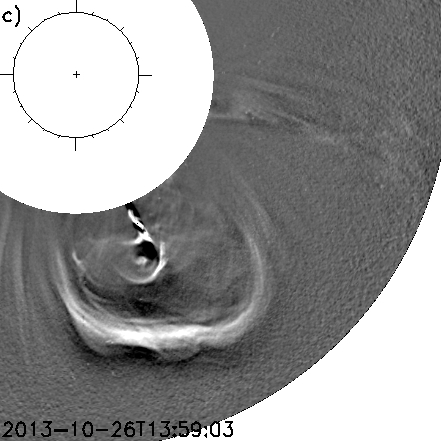}
\caption{2013-10-26 Eruption. a) and b) show the MGN processed images from SDO-AIA of the event with no clear loop structures overlaying the erupting filament. c) Dynamic C2 image of the eruption which has a clear 3-part CME structure, despite the lack of corresponding loop structures in the processed AIA data.}
\label{fig:13}
\end{figure*}

Category 2 consists of CMEs with clear 3-part structure, but are associated with erupting filaments without observable pre-existing loop structures. After the exclusion of the 88 failed eruptions, this is the largest category, containing 61 out of the main 133 events. As an example, the filament eruption observed by AIA on 2013/10/26 11:00 near the South-West limb is seen in figure \ref{fig:13}a,b to have no loop structure overlaying the filament prior to or during the eruption. However, as shown in figure \ref{fig:13}c, the corresponding CME is seen to have a clear 3-part structure in LASCO-C2, with the core of the ejection easily being identified as the filament eruption as much of the morphology is maintained. The search for pre-existing loops was repeated without success in hotter EUV wavelength channels (193\r{A} and 211\r{A}), and in the EUVI imagers of STEREO A and B. There can be several observational reasons why pre-existing loops may exist but are simply not observable, as listed in section \ref{sec:discuss}. In this case, we believe that if there is a pre-existing magnetic structure, the alignment of the pre-erupted filament is not favourable, and that the apex of the flux tube may be too high in the corona for detection.

\FPset\figsiz{5.0}

\subsection{Category 3}
\label{sec:cat3}
\begin{figure*}[t]
\centering
\includegraphics[width=\figsiz cm]{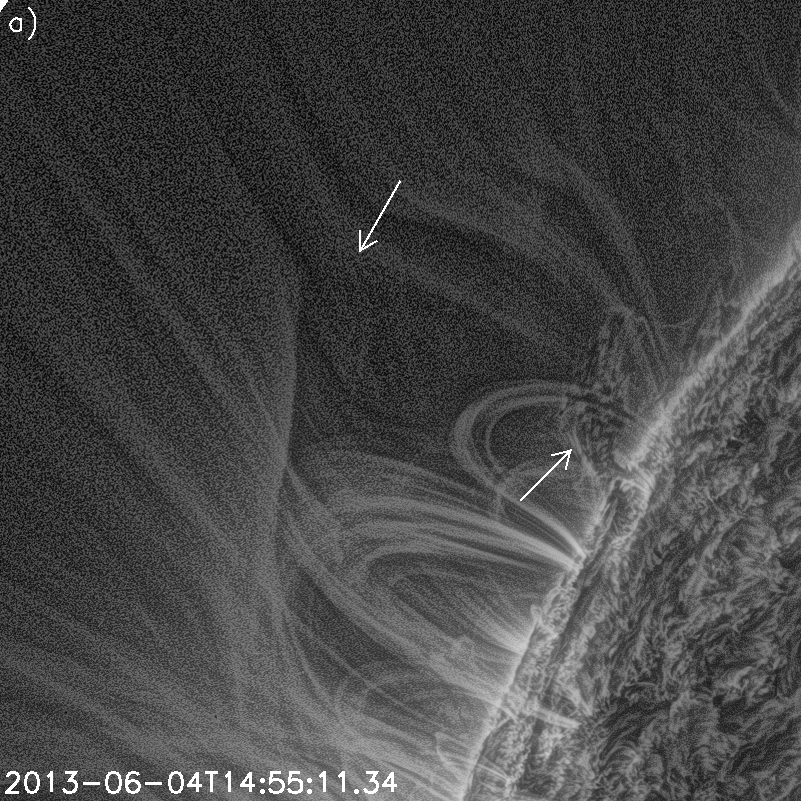}
\includegraphics[width=\figsiz cm]{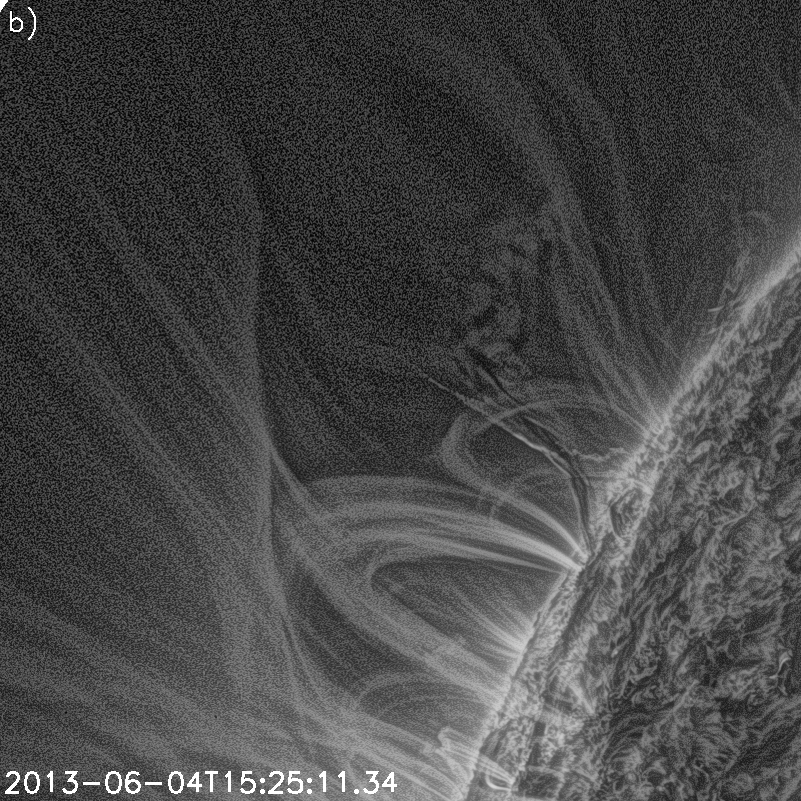}
\includegraphics[width=\figsiz cm]{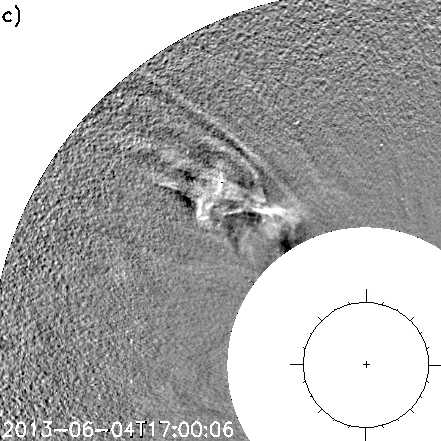}
\caption{2013-06-04 category 3 eruption. a) and b) show the MGN processed images from SDO-AIA of the erupting filament. c) Dynamic C2 image of the eruption which has no discernible structure, despite the loop structure seen in AIA.}
\label{fig:14}
\end{figure*}

Category 3 events are filaments that have pre-existing loop structures, but lead to either very faint disturbances in the extended corona, or CMEs which lack a clear 3-part structure. From the 133, only 2 events fall into this category. The 2013/06/04 14:00 eruption from the North-East limb of AIA is one example. Figure \ref{fig:14}a,b show EUV images with a clear loop structure overlaying the filament prior to, and which expands outwards with, the eruption. However the corresponding eruption seen later in the LASCO-C2 white light image of figure \ref{fig:14}c has no 3-part structure, and certainly no clear sign of a bright leading edge. This is an interesting case. The eruption is not from a compact active region, yet the overlying and surrounding magnetic structure seems very complicated. The overlying field seems not to be formed from large continuous loops, but instead contains a gap labelled by the white arrow in figure \ref{fig:14}a. This configuration is disrupted and moved aside by the erupting filament, but does not form an enclosing closed sheath. From a comparison of structure of the erupting filament and the white light CME, we suggest that this CME is formed solely of the material of the filament itself, and the magnetic structure may be of a compact flux tube containing the filament. That is, there is no extended flux tube which leads to a 3-part CME structure.

\FPset\figsiz{5.0}

\subsection{Category 4}
\label{sec:cat4}
\begin{figure*}[t]
\centering
\includegraphics[width=\figsiz cm]{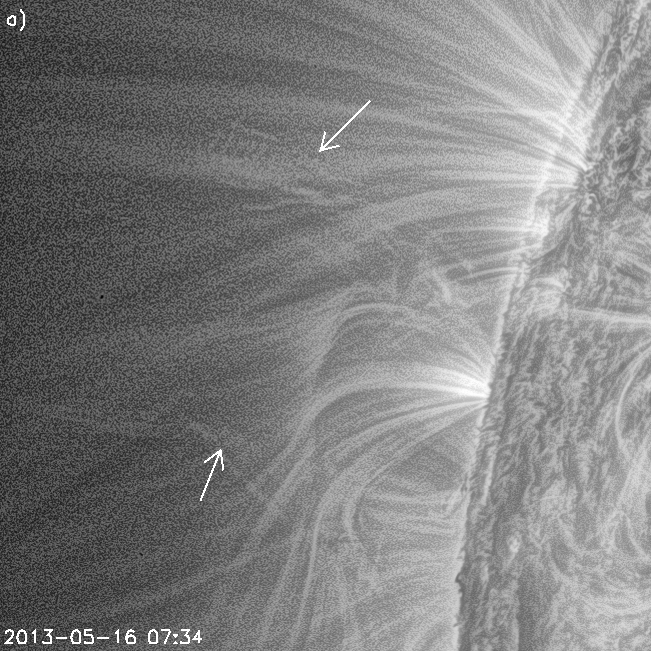}
\includegraphics[width=\figsiz cm]{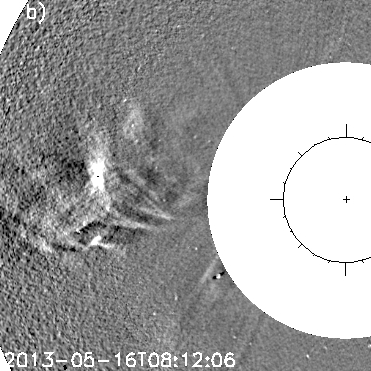}
\caption{2013-05-16 category 4 event. Eruption of a filament without pre-existing loop structures in EUV (a), and the accompanying CME which has no clear structure (b).}
\label{fig:cat4}
\end{figure*}

This category consists of 39 CMEs or faint 'blobs' visible in the LASCO C2 FOV without clear 3-part structure, associated with filament eruptions which show no sign of overlying loop structures. Figure \ref{fig:cat4} shows an example of such an unstructured CME in the LASCO C2 field of view and the associated filament eruption.

\FPset\figsiz{5.0}

\section{Discussion}
\label{sec:discuss}

Given the careful selection of the 31 Category 1 events, it is clear that the loop structures which overlie filaments (i.e. cavities) can erupt with the filament and, furthermore, form the bright leading edges of CMEs in the extended corona. Despite the data gap between heights of $\sim$0.3 and 1.2\Rs\ above the limb, the argument is convincing, particularly given the good agreement between a flux rope density model and observation, with the kinematics and position of the eruption used as input for the model. This supports the model of filament/cavity systems as two observable features (the filament and overlying loops) of a single entity (large magnetic flux tube, with the filament lying at the bottom of the tube). Just over a third of all events with clear 3-part CMEs are category 1 events. 

Category 3 and 4 events are CMEs with no clear 3-part structure as observed in coronagraph images. There are a total of 41 events belonging to these categories, of which only 2 events started with eruptions which had a clear set of pre-existing loops. Category 1 and 2 events are CMEs with a clear 3-part structure, of which 31 have clear pre-existing loop structures which form the leading part of the CME, giving a ratio of $\sim$34\%. CMEs with a clear 3-part structure are therefore far more likely to arise from eruptions which have a front edge formed from a pre-existing magnetic structure in the low corona. 

Around 66\%\ of clear 3-part CMEs as viewed in coronagraph images arise from erupting filaments without clear evidence of a system of pre-existing and co-erupting loops (category 2). There are several reasons that pre-existing loops may exist for these events, but are simply not observed. The first three reasons listed here are likely the most important:
\begin{itemize}
\item The pre-existing cavity loops, if present,  lie at a height either beyond the field of view of the EUV imager, or at a height where signal is too low to detect even using advanced image processing. 
\item The pre-existing cavity loops, if present, are obscured by other quiescent structures along the line of sight. This is particularly relevant to active region eruptions.
\item The pre-existing cavity loops, if present, are not aligned favourably relative to the observer. A favourable alignment would be East-West. If the cavity structure is misaligned from East-West, it becomes more difficult to detect due to the line-of-sight integration.
\item The temperature of the pre-existing cavity loops may not fall favourably into the AIA channels, although our search was made in the 171, 193 and 211\AA\ channels which does give a large appropriate temperature range. 
\item The magnetic cavity/loop structure may be at a density which is not significantly above the background medium, that is, the magnetic structure is not appreciably traced by the density structure. 
\end{itemize}
It is likely that these factors, particularly the first three observational reasons, lead to a large number of Category 2 events which have pre-existing loops which are simply not observable. The 33\%\ of 3-part CMEs which arise from erupting filaments which have observed systems of enclosing loops (i.e. cavities) is therefore very much a lower limit. Many of the category 2 events probably have pre-existing systems of enclosing loops which are simply not observed or detectable in EUV images.

\begin{figure}[t]
\centering
\includegraphics[scale= 0.8]{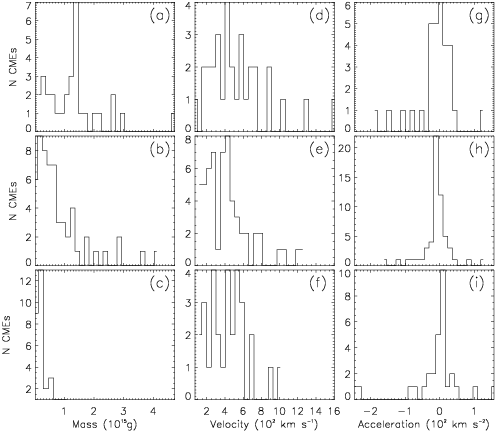}
\caption{Histograms of CME mass (left column), velocity (middle column) and acceleration (right column) for category 1 (top row), category 2 (middle row) and category 4 events (bottom row).}
\label{fig:histograms}
\end{figure}

For all events, the velocity, acceleration and mass of each CME was estimated, as listed in the table in the appendix. The velocities and accelerations were estimated using the bootstrapping method. As all events erupted from close to the limb, no correction has been made for projection effects. The mass of each CME was calculated using the DST for isolating the CME signal in LASCO C2 data, and assuming that the measured CME intensity all arose from electrons situated in the point of closest approach along each line of sight (plane-of-sky approximation). Histograms of mass, velocity and acceleration are plotted for category 1, 2 and 4 events in figure \ref{fig:histograms}. There is an obvious difference between the mass of non-structured CMEs (category 4) and the other two categories. These CMEs are considerably less massive than the structured CMEs. There are no significant differences in the distribution of velocities and acceleration between all three categories. Category 1 events are peaked in mass at $\sim1.4\times10^{15}g$, whilst the category 2 events peak at a lower mass, but with a good number of CMEs at higher masses. There is likely a connection between the higher most probable mass of category 1 events compared to category 2. Category 1 events all have pre-existing, detectable, systems of surrounding loops which erupt with the filament. The surrounding loop systems exist at heights which can be detected by EUV imagers (i.e. the higher-density lower corona), and/or are bright enough to be detectable. It is not surprising therefore to find that the associated CMEs in coronagraph data are more likely to be brighter. A similar argument holds for projection effects. If the surrounding loop systems are more easily detected in category 1 events due to favorable alignment, they will likely appear brighter in the coronagraph data.

By a manual `point-and-click' method, the core regions of the 3-part structured CMEs are isolated from the rest of the CME, and a mass calculated separately for the core. This is applied to the category 1 and 2 events. The mean core mass is $1.18\times10^{15}g$ for category 1 and $8.5\times10^{14}g$ for category 2. The mean (total) mass for category 4 unstructured CMEs is $3.9\times10^{14}g$. The mean core mass to total mass ratio is 0.26 for category 1, and 0.34 for category 2. All these values have a large variation. In general, and given the large uncertainties inherent in CME mass estimates, the masses of category 4 unstructured CMEs are of the same order of magnitude as the cores of 3-part structured CMEs. The bulk of the 3-part CME mass is contained outside the core, and it seems as if non-3-part-CMEs do not possess this additional mass or structure.

%\FPset\figsiz{0.2}
%\begin{figure}[t]
%\label{fig:histograms}
%\centering
%\includegraphics[scale= \figsiz]{histograms/FR_filament_vel.png}
%\includegraphics[scale= \figsiz]{histograms/FR_filament_acc.png}
%\includegraphics[scale= \figsiz]{histograms/FR_filament_mass.png}
%\caption{{\small Distribution of flux-rope CME velocity (top), acceleration (middle) and mass (bottom) for those with loop structures overlaying the associated filament eruption (red) and those without (black).}}
%\end{figure}

Table \ref{tab:position} lists category 1 and 2 events according to region at the limb. The argument that some pre-existing loops may not be detected due to alignment effects can be strengthened by a significant difference between east and west limbs. Not including the north and south regions, there are 49 clear 3-part structure CMEs in the east, and 40 in the west. Of these, 27\% have pre-existing loops in the east compared to 43\% in the west. All regions show a similar ratio of category 1 events to category 1 and 2 events, except for the north-west which has an almost equal number in each category. In truth, our numbers are too small to form any conclusions from this information and an extended study is required. Furthermore, we would expect a relationship to become apparent if the eruption of extended filaments only were considered. For small filaments or active region eruptions, such a dependence of region on alignment is not expected.

\begin{table}[t]
\centering
\caption{Distribution of Category 1 and 2 events (as defined in Section \ref{sec:results}) according to region of the limb. The total column gives the total number of CMEs (all events).}
\begin{tabular}{l|r|r|r|r|r}
\hline\hline
Region & Total & Non-fails & Cat. 1 \& 2 & Cat. 1 & Ratio
\\\hline
S  & 6 & 4 & 3 & 1 & 0.333\\
SE & 46 & 32 & 23 & 5 & 0.217\\
E  & 27 & 18 & 10 & 3 & 0.3\\
NE & 32 & 21 & 16 & 5 & 0.313\\
N  & 7 & 1 & 0 & 0 & 0\\
NW & 37 & 17 & 12 & 7 & 0.583\\
W  & 28 & 15 & 11 & 3 & 0.273\\
SW & 38 & 25 & 17 & 7 & 0.412
\\\hline
\end{tabular}
\label{tab:position}
\end{table}

Category 3 is a small yet interesting set of 2 events. These are filaments with clear pre-existing loops which erupt along with the filament, but which fail to form clear 3-part CMEs in coronagraph images. The case event from this category (2013-06-04 event described in section \ref{sec:cat3}) erupted to form a ejection seen by LASCO/C2, but which had no clear structure, despite the clear loop structures seen by AIA. The erupting filament has only a very small surrounding loop structure, and this structure seems to lie under a larger system of complicated field which may not be entirely closed (see figure \ref{fig:14}). In truth, this one case may well belong in category 4. The large outmost overlying magnetic structure (which is draping over the erupting filament and other closed systems of loops), does not appear to be a closed loop. The resulting CME therefore does not possess an enclosing large expanding flux tube, and is composed of just a core which may be a compact magnetic flux tube with a complicated structure.
88 events were discarded from the survey as they had no CME because the eruption failed to reach the coronagraph field of view. Within this set, 3 events are seen to have pre-existing loop systems. The loop systems seen with these filaments expand outwards with the initial expansion, but for unknown reasons, fail to form a CME.

The emphasis of our study has been on the detection of pre-existing loops. Category 1 and 2 events formed structured 3-part CMEs, and the high percentage (given the observational limits) of category 1 events, suggests that all 3-part CMEs are formed from the eruption of pre-existing large flux tubes in the low corona. This is consistent with current models of filament/cavity systems, CME initiation and 3-part CMEs in the extended corona and heliosphere. 39 of the events (category 4) did not possess observable pre-existing loops, and did not form structured CMEs. Furthermore, the unstructured CMEs were considerably less massive than the 3-part structured CMEs, and their masses are similar to the masses of 3-part CME cores. An important question is what is the reason for their lower mass and unstructured appearance? Prior to eruption, if they exist as extended flux tubes, their outer envelopes must be disrupted by some process during eruption in the low corona, which allows only the core filament eruption to escape. Our study strongly suggests this is not the case since we do not find any evidence of pre-existing loops which would be evidence of a large flux tube. We suggest that they exist initially as compact flux tubes which forms the dense filament. These compact structures can erupt and do not appear as clear 3-part CMEs in the extended corona since they do not possess a large flux tube which encloses the underlying filament. If this is true, then there is a large family of eruptive filaments which are not associated with a large overlying flux tube and do not fit into the model of a large erupting flux tube. Rather, they are narrow, constrained flux tubes containing only the filament material. Their signature, if measurable, in {\it in situ} measurements should be distinct from the magnetic clouds arising from larger structured CMEs.

In the context of eruptions which form higher in the corona (at the cusp of helmet streamers), \citet{sheeley2009} have shown that unstructured `blob' CMEs as seen in coronagraphs are consistent with narrow extended flux tubes. Category 4 events are different, being formed from filament eruptions near the Sun. Jetting activity in, or near, active regions can also lead to `puff' CMEs which lack internal structure. \citet{Yu_2014} show such jet CMEs propagating to large distances from the Sun and into interplanetary space. A recent study of a rapid series of 12 jet CMEs arising from a single region over a $\sim$2-day period was made by \citet{alzate2015}, giving mass estimates in the range of 0.4-3.2$\times10^{14}g$, with a mean mass of 1.8$\times10^{14}g$. These are considerably less massive than even the category 4 events. The category 4 events lack a clear 3-part structure, but they do contain internal structure, albeit complicated and difficult to interpret. The jet CMEs presented by \citet{alzate2015} lack internal structure, they appear as unstructured `puffs' of material.

Our study has not made an attempt to categorize filament eruptions according to their association with active regions. Outside of active regions, large filaments which are embedded within a system of enclosing loops are commonly filament-cavity systems. Within active regions, filaments cannot be associated with cavity strutures. Nevertheless, many of the active region erupting filaments are seen to possess large systems of loops which erupt with the filament. These are extended canopies of closed fields or arcades within which the filament is embedded, and are seen to erupt with the filament (see for example the Category 1 event of 2013/08/27). 

As a last point of discussion, the loop structures seen in EUV images, lying quiescently in the low corona long prior to the eruption, are often very difficult to observe. Advanced image processing, such as the MGN process used in this work, is cruicial to reveal such faint structures. Time differencing, often used for studies of eruptions, would not help in such a study. To the contrary, time differencing would conceal any evidence of a pre-existing static structure, and its eventual eruption would manifest as a sudden appearance of material at the front of the CME. This is of course an artifact of the process and highlights how time differencing can lead to misinterpretation of coronal data. 

\section{Conclusion}
\label{sec:conclusion}

221 solar filament eruptions which erupted very near or on the limb as observed from the Sun-Earth line are studied in detail in EUV images of the low corona, and in white light coronagraph images of the extended inner corona. From this study, we find that:
\begin{itemize}
\item $\sim$40\%\ were failed eruptions (no CMEs). Of the remaining 133 filament eruptions, 94 led to 3-part structured CMEs (70\%), and 41 led to unstructured CMEs (30\%).
\item 3-part CMEs are large flux tubes, with the leading edge of the CME corresponding to the outer envelope of the flux tube, and the inner core of the CME corresponding to the dense filament material near the base of the flux tube.
\item Around 33\%\ of 3-part CMEs arise from filament/enclosing loop systems, where the enclosing loops erupt with the filament and forms the leading bright edge of the CME. Due to observation limitations, the 33\%\ is very much a lower limit. In the case of non-active region filaments, the enclosing loop systems are cavities.
\item In general, brighter 3-part CMEs are more likely to arise from erupting filaments which possess a detectable system of enclosing loops. This can be interpreted as an observational bias (brighter CMEs $=$ detectable loops).
\item Only 2 of the 41 unstructured CMEs possess pre-existing systems of loops, and these 2 events are suspect in that there is some ambiguity in the definition of their pre-existing loops. 
\item Unstructured CMEs are significantly less bright/massive than the 3-part CMEs. Their mass is of the same order of magnitude as the masses of 3-part CME cores. The bulk of 3-part CME mass is outside of the core region.
\end{itemize}

We conclude that the bright frontal structures of 3-part CMEs arise from pre-existing structures in the low corona - namely the systems of closed loops which surround filaments. This bright front edge is not the shock wave which accompanies many active region impulsive eruptions - it is part of the magnetic structure of the eruption and expands behind any shock wave. The study of category 1 and 2 events are consistent with two connected models: 1) filament/cavity systems as two observable parts of one structure - a large magnetic flux tube embedded in the low corona, with the dense filament material lying at the bottom of the flux tube, and 2) structured 3-part CMEs as the subsequent eruption and expansion of the large flux tube, with the core formed from dense filament material distributed along the bottom of the tube.

This study also suggests that non-3-part CMEs are a different family to 3-part CMEs. They do not arise from the eruption of large filament/enclosing flux tube (cavity) systems, but rather from the eruption of a filament in the absence of a larger enclosing flux tube. We believe them to be compact flux tubes containing only the dense filament material. This has implications for models of filaments in general (in the absence of eruption), models of CME initiation and eruption, and for the detection and interpretation of ICMEs in interplanetary measurements.

A recent study by \citet{Wang15} looked carefully at the overlaying magnetic field configuration of pseudostreamers as cause for lower-mass unstructured CMEs using potential-field source-surface extrapolation. For a future study, we propose to apply the EUV image processing used in this paper on pseudostreamer source events to establish more clearly the link between CME structure and the structure of the erupting filament system. Whether or not the results of this research match those found here could be a very important finding.

\subparagraph{Acknowledgements}
Joe's work is conducted under an STFC studentship to the Solar Systems Physics group at Aberyswyth University. Huw is grateful for a research fellowship from the Leverhulme Foundation, which made this work possible. The SOHO/LASCO data used here are produced by a consortium of the Naval Research Laboratory (USA), Max-Planck Insitut f{\"u}r Aeronomie (Germany), Laboratoire d'Astronomie (France), and the University of Birmingham (UK). SOHO is a project of international cooperation between ESA and NASA. The STEREO/SECCHI project is an international consortium of the Naval Research Laboratory (USA), Lockheed Martin Solar and Astrophysical Laboratory (USA), NASA Goddard Space Flight Center (USA), Rutherford Appleton Laboratory (UK), University of Birmingham (UK), Max-Planck Institut f{\"u}r Sonnen-systemforschung (Germany), Centre Spatial de Liege (Belgium), Institut d'Optique Th{\'e}orique et Appliqu{\'e}e (France), and Institut d'Astrophysique Spatiale (France). The SDO data used are provided courtesy of NASA/SDO and the AIA science team.

%\bibliography{adsbib.bib}

\begin{thebibliography}{33}

\bibitem[Alzate \& Morgan(2015)]{alzate2015} Alzate, N. \& Morgan, H. \ 2015, ApJ, submitted (06/2015)

\bibitem[{{Antiochos} {et~al.}(1999){Antiochos}, {DeVore}, \&
  {Klimchuk}}]{antiochos99}
{Antiochos}, S.~K., {DeVore}, C.~R., \& {Klimchuk}, J.~A. 1999, ApJ, 510, 485

\bibitem[{{Berger} {et~al.}(2012){Berger}, {Liu}, \& {Low}}]{berger2012}
{Berger}, T.~E., {Liu}, W., \& {Low}, B.~C. 2012, ApJ letters, 758, L37

\bibitem[{{Brueckner} {et~al.}(1995){Brueckner}, {Howard}, {Koomen},
  {Korendyke}, {Michels}, {Moses}, {Socker}, {Dere}, {Lamy}, {Llebaria},
  {Bout}, {Schwenn}, {Simnett}, {Bedford}, \& {Eyles}}]{brueckner95}
{Brueckner}, G.~E., {et~al.} 1995, SolPhys, 162, 357

\bibitem[{{Byrne} {et~al.}(2013){Byrne}, {Long}, {Gallagher}, {Bloomfield},
  {Maloney}, {McAteer}, {Morgan}, \& {Habbal}}]{byrne13}
{Byrne}, J.~P., {Long}, D.~M., {Gallagher}, P.~T., {Bloomfield}, D.~S.,
  {Maloney}, S.~A., {McAteer}, R.~T.~J., {Morgan}, H., \& {Habbal}, S.~R. 2013, A\&A, 557, A96

\bibitem[{{Chen}(2011)}]{chen11}
{Chen}, P.~F. 2011, Living Reviews in Solar Physics, 8, 1

\bibitem[{{Chen} \& {Shibata}(2000)}]{chen00}
{Chen}, P.~F., \& {Shibata}, K. 2000, ApJ, 545, 524

\bibitem[{{Cheng} {et~al.}(2014){Cheng}, {Ding}, {Guo}, {Zhang}, {Vourlidas},
  {Liu}, {Olmedo}, {Sun}, \& {Li}}]{cheng2014}
{Cheng}, X., {et~al.} 2014, ApJ, 780, 28

\bibitem[Colaninno 
\& Howard(2015)]{colaninno2015} Colaninno, R.~C., \& Howard, R.~A.\ 2015, SolPhys, 290, 997 

\bibitem[{{Efron}(1979)}]{efron79}
{Efron}, B. 1979, The Annals of Statistics, 7

\bibitem[{{Gibson} {et~al.}(2010){Gibson}, {Kucera}, {Rastawicki}, {Dove}, {de
  Toma}, {Hao}, {Hill}, {Hudson}, {Marqu{\'e}}, {McIntosh}, {Rachmeler},
  {Reeves}, {Schmieder}, {Schmit}, {Seaton}, {Sterling}, {Tripathi},
  {Williams}, \& {Zhang}}]{gibson2010}
{Gibson}, S.~E., {et~al.} 2010, ApJ, 724, 1133

\bibitem[Gopalswamy(2006)]{gop2006} Gopalswamy, N.\ 2006, SSRv, 
124, 145 

\bibitem[{{Habbal} {et~al.}(2010){Habbal}, {Druckm{\"u}ller}, {Morgan},
  {Scholl}, {Ru{\v s}in}, {Daw}, {Johnson}, \& {Arndt}}]{habbal2010}
{Habbal}, S.~R., {Druckm{\"u}ller}, M., {Morgan}, H., {Scholl}, I., {Ru{\v
  s}in}, V., {Daw}, A., {Johnson}, J., \& {Arndt}, M. 2010, ApJ, 719, 1362

\bibitem[Habbal et al.(2013)]{habbal2013} Habbal, S.~R., Morgan, 
H., Druckm{\"u}ller, M., et al.\ 2013, SolPhys, 285, 9 

\bibitem[{{Habbal} {et~al.}(2014){Habbal}, {Morgan}, \&
  {Druckm{\"u}ller}}]{habbal2014}
{Habbal}, S.~R., {Morgan}, H., \& {Druckm{\"u}ller}, M. 2014, ApJ, 793, 119

\bibitem[Halain et al.(2013)]{halain2013} Halain, J.-P., 
Berghmans, D., Seaton, D.~B., et al.\ 2013, SolPhys, 286, 67 

\bibitem[{{Howard} {et~al.}(2008){Howard}, {Moses}, {Vourlidas}, {Newmark},
  {Socker}, {Plunkett}, {Korendyke}, {Cook}, {Hurley}, {Davila}, {Thompson},
  {St Cyr}, {Mentzell}, {Mehalick}, {Lemen}, {Wuelser}, {Duncan}, {Tarbell},
  {Wolfson}, {Moore}, {Harrison}, {Waltham}, {Lang}, {Davis}, {Eyles},
  {Mapson-Menard}, {Simnett}, {Halain}, {Defise}, {Mazy}, {Rochus}, {Mercier},
  {Ravet}, {Delmotte}, {Auchere}, {Delaboudiniere}, {Bothmer}, {Deutsch},
  {Wang}, {Rich}, {Cooper}, {Stephens}, {Maahs}, {Baugh}, {McMullin}, \&
  {Carter}}]{howard08}
{Howard}, R.~A., {et~al.} 2008, SSRv, 136, 67

\bibitem[Jackson 
\& Hildner(1978)]{jackson1978} Jackson, B.~V., \& Hildner, E.\ 1978, SolPhys, 60, 155 

\bibitem[{{Jing} {et~al.}(2004){Jing}, {Yurchyshyn}, {Yang}, {Xu}, \&
  {Wang}}]{jing04}
{Jing}, J., {Yurchyshyn}, V.~B., {Yang}, G., {Xu}, Y., \& {Wang}, H. 2004, ApJ, 614, 1054

\bibitem[\protect\citeauthoryear{Labrosse et 
al.}{2010}]{labrosse2010} Labrosse N., Heinzel P., Vial J.-C., 
Kucera T., Parenti S., Gun{\'a}r S., Schmieder B., Kilper G., 2010, SSRv, 
151, 243 

\bibitem[{{Landi} {et~al.}(2010){Landi}, {Raymond}, {Miralles}, \&
  {Hara}}]{landi2010}
{Landi}, E., {Raymond}, J.~C., {Miralles}, M.~P., \& {Hara}, H. 2010, ApJ,  711, 75

\bibitem[{{Lemen}(2014)}]{lemen14}
{Lemen}, J. 2014, American Astronomical Society Meeting Abstracts, Vol. 223, 460.01

\bibitem[Lepri 
\& Zurbuchen(2010)]{lepri2010} Lepri, S.~T., \& Zurbuchen, T.~H.\ 2010, ApJ letters, 723, L22 

\bibitem[{Liu {et~al.}(2010)Liu, Thernisien, Luhmann, Vourlidas, Davies, Lin,
  \& Bale}]{lui10}
Liu, Y., Thernisien, A., Luhmann, J.~G., Vourlidas, A., Davies, J.~A., Lin,
  R.~P., \& Bale, S.~D. 2010, ApJ, 722, 1762

\bibitem[Liu et al.(2012)]{liu2012} Liu, W., Berger, T.~E., 
\& Low, B.~C.\ 2012, ApJ letters, 745, L21 

\bibitem[{{Low} {et~al.}(2012{\natexlab{a}}){Low}, {Berger}, {Casini}, \&
  {Liu}}]{low2012b}
{Low}, B.~C., {Berger}, T., {Casini}, R., \& {Liu}, W. 2012{\natexlab{a}}, ApJ, 755, 34

\bibitem[{{Low} {et~al.}(2012{\natexlab{b}}){Low}, {Liu}, {Berger}, \&
  {Casini}}]{low2012a}
{Low}, B.~C., {Liu}, W., {Berger}, T., \& {Casini}, R. 2012{\natexlab{b}}, ApJ, 757, 21

\bibitem[\protect\citeauthoryear{Mackay et al.}{2010}]{mackay2010} 
Mackay D.~H., Karpen J.~T., Ballester J.~L., Schmieder B., Aulanier G., 
2010, SSRv, 151, 333 

\bibitem[{{Moore} \& {Roumeliotis}(1992)}]{moore92}
{Moore}, R.~L., \& {Roumeliotis}, G. 1992, in Lecture Notes in Physics, Berlin
  Springer Verlag, Vol. 399, IAU Colloq. 133: Eruptive Solar Flares, ed.
  Z.~{Svestka}, B.~V. {Jackson}, \& M.~E. {Machado}, 69

\bibitem[{{Morgan}(2010)}]{morgan10}
{Morgan}, H. 2010, in Bulletin of the American Astronomical Society, Vol.~41,
  American Astronomical Society Meeting Abstracts, 216, 314.01

\bibitem[{{Morgan}(2015)}]{morgan2015}
{Morgan}, H. 2015, ApJ, accepted for publication (06/2015)

\bibitem[{{Morgan} {et~al.}(2012){Morgan}, {Byrne}, \& {Habbal}}]{auto1}
{Morgan}, H., {Byrne}, J.~P., \& {Habbal}, S.~R. 2012, ApJ, 752, 144

\bibitem[{{Morgan} \& {Druckm{\"u}ller}(2014)}]{mgn}
{Morgan}, H., \& {Druckm{\"u}ller}, M. 2014, SolPhys, 289, 2945

\bibitem[{{Pagano} {et~al.}(2014){Pagano}, {Mackay}, \& {Poedts}}]{pagano14}
{Pagano}, P., {Mackay}, D.~H., \& {Poedts}, S. 2014, A\&A, 568, A120

\bibitem[{{Panesar} {et~al.}(2014){Panesar}, {Innes}, {Schmit}, \&
  {Tiwari}}]{panesar14}
{Panesar}, N.~K., {Innes}, D.~E., {Schmit}, D.~J., \& {Tiwari}, S.~K. 2014, SolPhys, 289, 2971

\bibitem[{{Qu{\'e}merais} \& {Lamy}(2002)}]{quemerais2002}
{Qu{\'e}merais}, E., \& {Lamy}, P. 2002, A\&A, 393, 295

\bibitem[{{Schwenn} {et~al.}(2005){Schwenn}, {dal Lago}, {Huttunen}, \&
  {Gonzalez}}]{schwenn05}
{Schwenn}, R., {dal Lago}, A., {Huttunen}, E., \& {Gonzalez}, W.~D. 2005,
  Annales Geophysicae, 23, 1033

\bibitem[Seaton et al.(2013)]{seaton2013} Seaton, D.~B., 
Berghmans, D., Nicula, B., et al.\ 2013, SolPhys, 286, 43 

\bibitem[{{Sheeley} {et~al.}(2009){Sheeley}, {Lee}, {Casto}, {Wang}, \&
  {Rich}}]{sheeley2009}
{Sheeley}, Jr., N.~R., {Lee}, D.~D.-H., {Casto}, K.~P., {Wang}, Y.-M., \&
  {Rich}, N.~B. 2009, ApJ, 694, 1471

\bibitem[{{Sterling} {et~al.}(2007){Sterling}, {Harra}, \&
  {Moore}}]{sterling07}
{Sterling}, A.~C., {Harra}, L.~K., \& {Moore}, R.~L. 2007, ApJ, 669, 1359

\bibitem[{{Thernisien} {et~al.}(2009){Thernisien}, {Vourlidas}, \&
  {Howard}}]{thernisien09}
{Thernisien}, A., {Vourlidas}, A., \& {Howard}, R.~A. 2009, SolPhys, 256, 111

\bibitem[{{T{\"o}r{\"o}k} \& {Kliem}(2005)}]{torok05}
{T{\"o}r{\"o}k}, T., \& {Kliem}, B. 2005, ApJ letter, 630, L97

\bibitem[{{Vourlidas} {et~al.}(2013){Vourlidas}, {Lynch}, {Howard}, \&
  {Li}}]{vourlidas2013}
{Vourlidas}, A., {Lynch}, B.~J., {Howard}, R.~A., \& {Li}, Y. 2013, SolPhys 284, 179
  
\bibitem[{Wang}(2015)]{Wang15}
{Wang},Y.M. 2015, ApJ 803, L12

\bibitem[{{Yashiro} {et~al.}(2004){Yashiro}, {Gopalswamy}, {Michalek},
  {St.~Cyr}, {Plunkett}, {Rich}, \& {Howard}}]{yashiro04}
{Yashiro}, S., {Gopalswamy}, N., {Michalek}, G., {St.~Cyr}, O.~C., {Plunkett},
  S.~P., {Rich}, N.~B., \& {Howard}, R.~A. 2004, Journal of Geophysical
  Research (Space Physics), 109, 7105

\bibitem[Yu et al.~(2014)]{Yu_2014} Yu, H.-S., Jackson, B. V., Buffington, A., et al., 2014, ApJ, 784, 166

\end{thebibliography}

\section{Appendix 1}
\label{sec:appendix}
A list of all events analysed in this study. Supplementary movies of the Figure 5, 6 and 7 events are available at:

\texttt{http:/users.aber.ac.uk/joh9/erupting$\_$filaments.html}

\begin{table}[htp]
\caption{List of the 221 filament eruptions from May 2013 to June 2014 examined. Using reduced chi squared, the uncertainties in velocity and acceleration were found to be 9.6 km.$s^{-1}$ and 4.5 m.$s^{-2}$ respectively.}
\centering
\footnotesize\setlength{\tabcolsep}{4pt}
\begin{tabular}{lrrlccrrr}

\hline \hline
Start Date/Time & Long & Lat & Region & 3-part CME & AIA loops & Mass (g) & v (km.$s^{-1}$) & a (m.$s^{-2}$)
\\\hline
2013-05-03T06:20:23 & -89 & 07 & E & N & Y &  &  &  \\
2013-05-06T01:00:03 & -78 & -19 & SE & N & N &  &  &  \\
2013-05-06T07:30:35 & -49 & -12 & SE & N & N &  &  &  \\
2013-05-07T05:00:03 & 78 & -75 & S & Y & Y & 3.83$\times$ $10^{14}$ & 231 & 13 \\
2013-05-10T00:00:03 & 85 & 56 & NW & N & N &  &  &  \\
2013-05-10T18:10:03 & 88 & 27 & NW & Y & N & 1.41$\times$ $10^{15}$ & 493 & 0 \\
2013-05-11T20:00:03 & 85 & 55 & NW & Y & N & 2.93$\times$ $10^{14}$ & 445 & 15 \\
2013-05-14T04:40:03 & -90 & -7 & E & N & N &  &  &  \\
2013-05-14T06:30:03 & 86 & -61 & SW & N & N &  &  &  \\
2013-05-14T21:10:03 & 90 & -11 & W & Y & Y & 8.72$\times$ $10^{14}$ & 676 & 4 \\
2013-05-15T01:40:03 & 88 & -47 & SW & N & N &  &  &  \\
2013-05-16T04:50:03 & -89 & -24 & SE & N & N &  &  &  \\
2013-05-16T07:10:03 & -89 & 10 & E & N & N & 1.09$\times$ $10^{15}$ & 953 & -41 \\
2013-05-18T01:04:03 & -89 & 10 & E & Y & Y & 3.40$\times$ $10^{14}$ & 68 & 7 \\
2013-05-18T07:02:03 & -89 & -21 & SE & N & N & 3.40$\times$ $10^{14}$ & 118 & 17 \\
2013-05-21T12:00:03 & 88 & 40 & NW & Y & Y & 2.33$\times$ $10^{14}$ & 443 & 18 \\
2013-05-22T08:00:35 & -88 & -47 & SE & N & N &  &  &  \\
2013-05-23T09:50:23 & -90 & -05 & E & N & N & 1.39$\times$ $10^{14}$ & 746 & 11 \\
2013-05-26T18:16:03 & -88 & 57 & NE & Y & N & 2.22$\times$ $10^{14}$ & 454 & -28 \\
2013-05-27T07:00:35 & 83 & 78 & N & N & N &  &  &  \\
2013-05-29T16:32:03 & 90 & -14 & W & N & N &  &  &  \\
2013-05-29T16:38:03 & 89 & 33 & NW & N & N &  &  &  \\
2013-05-30T08:06:03 & -89 & 27 & NE & Y & N & 1.33$\times$ $10^{15}$ & 494 & -34 \\
2013-05-31T05:48:03 & 90 & -24 & SW & Y & Y & 1.53$\times$ $10^{15}$ & 529 & 43 \\
2013-05-31T21:30:03 & 90 & -30 & SW & N & N &  &  &  \\
2013-06-04T13:00:03 & -89 & 46 & NE & N & Y & 3.58$\times$ $10^{14}$ & 683 & 98 \\
2013-06-09T08:00:03 & 90 & 55 & NW & N & N &  &  &  \\
2013-06-10T06:32:03 & 63 & 14 & W & Y & Y & 6.03$\times$ $10^{14}$ & 331 & -67 \\
2013-06-11T00:00:03 & 90 & 21 & NW & Y & N & 1.46$\times$ $10^{15}$ & 434 & -12 \\
2013-06-12T00:00:03 & 90 & 50 & NW & N & N &  &  &  \\
2013-06-12T10:48:03 & -69 & 89 & N & N & N &  &  &  \\
2013-06-13T11:20:03 & 90 & 32 & NW & N & N & 4.22$\times$ $10^{14}$ & 378 & -247 \\
2013-06-15T22:10:03 & 86 & -70 & SW & Y & N & 3.28$\times$ $10^{14}$ & 137 & 26 \\
2013-06-16T03:55:03 & 90 & 29 & NW & N & N & no data & 307 & 77 \\
2013-06-17T00:40:03 & 87 & -62 & SW & N & N &  &  &  \\
2013-06-17T00:40:03 & -90 & 32 & NE & N & N &  &  &  \\
2013-06-17T12:00:03 & -87 & -64 & SE & Y & N & 8.73$\times$ $10^{14}$ & 242 & 21 \\
2013-06-18T01:30:03 & 89 & 35 & NW & N & N & no data & 341 & -40 \\
2013-06-22T08:00:03 & -52 & 45 & NE & N & N &  &  &  \\
2013-06-30T18:00:03 & -88 & -25 & SE & N & N & 2.27$\times$ $10^{14}$ & 646 & 56 \\
2013-07-01T11:32:03 & -87 & -37 & SE & N & N &  &  &  \\
2013-07-02T18:27:35 & 47 & 86 & N & N & N &  &  &  \\
2013-07-06T18:24:03 & -88 & -23 & SE & Y & N & 1.27$\times$ $10^{15}$ & 506 & 15 \\
2013-07-13T05:40:03 & -88 & -17 & E & N & N &  &  &  \\
2013-07-17T20:24:03 & -89 & -03 & E & N & N & 4.74$\times$ $10^{14}$ & 294 & 10 \\
2013-07-18T16:30:03 & -89 & -11 & E & Y & Y & 2.77$\times$ $10^{15}$ & 393 & -16 \\
2013-07-18T19:39:03 & 88 & -17 & W & N & N &  &  &  \\
\hline\\
\end{tabular}
\end{table}
\begin{table}[htp]
\centering
\footnotesize\setlength{\tabcolsep}{4pt}
\begin{tabular}{lrrlccrrr}
\hline \hline
Start Date/Time & Long & Lat & Region & 3-part CME & AIA loops & Mass (g) & v (km.$s^{-1}$) & a (m.$s^{-2}$)
\\\hline
2013-07-21T20:30:03 & 83 & 56 & NW & Y & N & 6.76$\times$ $10^{14}$ & 320& 2 \\
2013-07-24T04:30:03 & 89 & 13 & W & N & N &  &  &  \\
2013-07-24T22:00:03 & 80 & 64 & NW & N & N &  &  &  \\
2013-07-29T22:10:03 & 89 & 11 & W & N & N &  &  &  \\
2013-07-31T17:06:03 & -86 & -32 & SE & N & N &  &  &  \\
2013-08-03T02:00:35 & 78 & 64 & NW & N & N &  &  &  \\
2013-08-04T06:15:35 & -56 & 24 & NE & Y & N & 2.23$\times$ $10^{14}$ & 403 & -6 \\
2013-08-04T14:30:35 & 55 & -20 & SW & N & N & 2.42$\times$ $10^{14}$ & 531 & 157 \\
2013-08-05T01:10:11 & 75 & -66 & SW & N & N &  &  &  \\
2013-08-05T13:50:23 & 88 & 20 & W & N & N &  &  &  \\
2013-08-06T01:12:03 & 88 & -19 & W & N & N & 2.25$\times$ $10^{14}$ & 712 & 36 \\
2013-08-06T01:16:03 & -50 & 28 & NE & Y & N & 6.94$\times$ $10^{14}$ & 612 & -29 \\
2013-08-06T05:28:03 & -86 & -29 & SE & N & N &  &  &  \\
2013-08-10T06:48:03 & 52 & -25 & SW & Y & N & 8.83$\times$ $10^{14}$ & 440 & 2 \\
2013-08-12T19:40:11 & -25 & -43 & SE & N & N &  &  &  \\
2013-08-14T05:00:03 & 81 & -51 & SW & N & N &  &  &  \\
2013-08-14T18:30:35 & 55 & -31 & SW & Y & N & 2.26$\times$ $10^{15}$ & 298 & 4 \\
2013-08-16T18:00:03 & -57 & 33 & NE & Y & N & 3.81$\times$ $10^{14}$ & 489 & 14 \\
2013-08-17T10:00:03 & -87 & 26 & NE & N & N &  &  &  \\
2013-08-17T16:00:03 & 60 & -33 & SW & Y & N & 4.19$\times$ $10^{15}$ & 1069 & 100 \\
2013-08-18T13:40:03 & 86 & -26 & SW & N & N & 3.75$\times$ $10^{13}$ & 501 & -7 \\
2013-08-18T19:44:03 & 85 & -32 & SW & N & N &  &  &  \\
2013-08-20T03:44:03 & 26 & -38 & S & Y & Y & 1.30$\times$ $10^{15}$ & 926 & 24 \\
2013-08-21T00:16:03 & 88 & 19 & W & N & N &  &  &  \\
2013-08-27T06:45:03 & 89 & -08 & W & Y & Y & 2.26$\times$ $10^{15}$ & 621 & -7 \\
2013-08-28T00:00:03 & 75 & 65 & NW & Y & N & 2.72$\times$ $10^{13}$ & 265 & 11 \\
2013-08-28T12:25:03 & 89 & -08 & W & N & N & 2.65$\times$ $10^{14}$ & 450 & 14 \\
2013-08-28T13:20:03 & -82 & -46 & SE & Y & N & 6.54$\times$ $10^{14}$ & 524 & 13 \\
2013-08-29T00:00:03 & 26 & -48 & SW & Y & N & 2.82$\times$ $10^{14}$ & 539 & 61 \\
2013-08-30T13:05:03 & -78 & -58 & SE & N & N & 3.49$\times$ $10^{14}$ & 324 & -8 \\
2013-08-31T00:00:03 & 84 & 43 & NW & N & N &  &  &  \\
2013-08-31T17:20:03 & -85 & -30 & SE & Y & N & 3.11$\times$ $10^{14}$ & 311 & 3 \\
2013-09-01T19:45:03 & -40 & -50 & SE & Y & N & 3.11$\times$ $10^{14}$ & 329 & 6 \\
2013-09-02T05:40:03 & -81 & 53 & NE & N & N &  &  &  \\
2013-09-04T20:00:03 & -63 & -16 & SE & N & N & 4.74$\times$ $10^{14}$ & 561 & 25 \\
2013-09-05T09:50:03 & 76 & 63 & NW & Y & N & 3.81$\times$ $10^{13}$ & 207 & 2 \\
2013-09-10T10:44:03 & -81 & -50 & SE & Y & Y & 7.19$\times$ $10^{14}$ & 748 & -38 \\
2013-09-11T08:00:03 & -82 & -45 & SE & Y & N & 6.17$\times$ $10^{14}$ & 412 & 39 \\
2013-09-13T14:00:03 & -20 & -82 & S & Y & N & 1.79$\times$ $10^{14}$ & 149 & 10 \\
2013-09-14T00:10:03 & -84 & -40 & SE & Y & N & 4.42$\times$ $10^{14}$ & 195 & 6 \\
2013-09-15T09:00:03 & -45 & -80 & SE & Y & N & 7.23$\times$ $10^{13}$ & 231 & 6 \\
2013-09-16T14:40:03 & 79 & 59 & NW & Y & N & 5.97$\times$ $10^{14}$ & 134 & 17 \\
2013-09-20T18:40:11 & -75 & 30 & NE & N & N &  &  &  \\
2013-09-20T20:00:03 & 82 & -46 & SW & N & N & 3.55$\times$ $10^{14}$ & 595 & 14 \\
2013-09-21T01:00:03 & -86 & 29 & NE & Y & N & 1.08$\times$ $10^{15}$ & 218 & 17 \\
2013-09-21T09:15:35 & -88 & -13 & E & N & N &  &  &  \\
2013-09-22T02:45:35 & 85 & -35 & SW & Y & N & 7.20$\times$ $10^{14}$ & 464 & 4 \\
2013-09-22T18:00:03 & 09 & -60 & S & N & N &  &  &  \\
2013-09-23T07:00:03 & 59 & -26 & SW & Y & N & 2.74$\times$ $10^{14}$ & 317 & -1 \\
2013-09-24T03:04:03 & -80 & -52 & SE & N & N & 1.73$\times$ $10^{14}$ & 168 & -29 \\
2013-09-24T18:36:03 & -86 & 31 & NE & Y & Y & 1.54$\times$ $10^{15}$ & 719 & 42 \\
2013-09-29T00:00:03 & 76 & 65 & NW & N & N &  &  &  \\
\hline\\
\end{tabular}
\end{table}
\begin{table}[htp]
\centering
\footnotesize\setlength{\tabcolsep}{4pt}
\begin{tabular}{lrrlccrrr}
\hline \hline
Start Date/Time & Long & Lat & Region & 3-part CME & AIA loops & Mass (g) & v (km.$s^{-1}$) & a (m.$s^{-2}$)
\\\hline
2013-10-01T14:00:03 & 77 & -61 & SW & Y & N & 2.95$\times$ $10^{15}$ & 239 & -14 \\
2013-10-02T23:10:03 & -50 & -14 & E & Y & N & 1.53$\times$ $10^{15}$ & 806 & 39 \\
2013-10-06T18:06:03 & -86 & -30 & SE & N & N &  &  &  \\
2013-10-07T06:57:35 & 82 & 52 & NW & N & N &  &  &  \\
2013-10-11T15:00:11 & 83 & 51 & NW & N & N &  &  &  \\
2013-10-12T13:00:03 & 88 & -18 & W & Y & Y & 1.53$\times$ $10^{15}$ & 357 & 4 \\
2013-10-12T16:00:03 & 86 & 39 & NW & N & N &  &  &  \\
2013-10-13T06:45:35 & -60 & -15 & E & N & N & 1.12$\times$ $10^{14}$ & 238 & 21 \\
2013-10-13T10:30:35 & 84 & -43 & SW & Y & N & 3.41$\times$ $10^{14}$ & 169 & 6 \\
2013-10-13T10:30:35 & 24 & 84 & N & N & N &  &  &  \\
2013-10-18T00:00:03 & -38 & 47 & NE & Y & N & 4.14$\times$ $10^{14}$ & 314 & 24 \\
2013-10-18T20:12:03 & 88 & -16 & W & N & N &  &  &  \\
2013-10-19T11:38:03 & -17 & 69 & N & N & N & 1.01$\times$ $10^{14}$ & 326 & 7 \\
2013-10-25T01:32:03 & -52 & 24 & NE & Y & N & 1.91$\times$ $10^{15}$ & 474 & 8 \\
2013-10-26T09:16:03 & 49 & -52 & SW & Y & N & 5.44$\times$ $10^{14}$ & 548 & 6 \\
2013-10-27T03:30:03 & -88 & -19 & E & N & N & no data & 57 & -39 \\
2013-10-28T02:42:03 & 88 & 22 & NW & Y & N & 1.34$\times$ $10^{15}$ & 686 & -160 \\
2013-10-28T10:28:03 & 42 & -52 & SW & N & N & 1.89$\times$ $10^{14}$ & 322 & 9 \\
2013-11-04T14:00:03 & 87 & 43 & NW & N & N &  &  &  \\
2013-11-04T18:20:23 & 59 & -55 & SW & N & N &  &  &  \\
2013-11-13T12:30:03 & -88 & 32 & NE & Y & N & 8.09$\times$ $10^{14}$ & 294 & 21 \\
2013-11-17T01:00:03 & 45 & -38 & SW & Y & Y & 9.45$\times$ $10^{14}$ & 707 & 38 \\
2013-11-17T09:00:03 & -51 & -38 & SE & Y & N & 1.20$\times$ $10^{15}$ & 396 & -3 \\
2013-11-22T13:36:03 & -89 & -09 & E & Y & Y & 4.62$\times$ $10^{14}$ & 181 & -3 \\
2013-12-08T03:12:03 & 90 & -02 & W & N & N &  &  &  \\
2013-12-09T10:32:03 & -90 & -37 & SE & N & N &  &  &  \\
2013-12-11T00:00:03 & -89 & 31 & NE & N & N &  &  &  \\
2013-12-11T05:30:03 & -59 & -58 & SE & Y & N & 1.32$\times$ $10^{14}$ & 340 & 4 \\
2013-12-12T17:00:03 & 89 & 50 & NW & N & N &  &  &  \\
2013-12-20T06:30:35 & 57 & 28 & NW & N & N &  &  &  \\
2013-12-20T20:00:31 & -88 & 40 & NE & Y & N & 5.91$\times$ $10^{13}$ & 118 & 3 \\
2013-12-21T03:20:23 & -90 & -22 & SE & N & N & 3.26$\times$ $10^{13}$ & 463 & 12 \\
2013-12-22T01:45:35 & -85 & 69 & NE & N & N &  &  &  \\
2013-12-22T04:30:35 & 88 & -48 & SW & N & N &  &  &  \\
2013-12-23T07:20:23 & 88 & 34 & NW & Y & N & 7.83$\times$ $10^{14}$ & 853 & 28 \\
2013-12-24T09:15:35 & -90 & -16 & E & N & N &  &  &  \\
2014-01-01T08:12:03 & -88 & -41 & SE & Y & N & 1.80$\times$ $10^{15}$ & 221 & 24 \\
2014-01-01T18:00:03 & -36 & -43 & SE & N & N &  &  &  \\
2014-01-03T00:40:03 & -88 & 23 & NE & N & N & 1.01$\times$ $10^{15}$ & 1026 & -233 \\
2014-01-05T02:20:03 & -90 & -01 & E & N & N & 3.47$\times$ $10^{14}$ & 629 & -80 \\
2014-01-06T05:10:03 & -88 & 31 & NE & N & N &  &  &  \\
2014-01-06T07:30:35 & 62 & -15 & W & Y & N & 3.75$\times$ $10^{15}$ & 1246 & -75 \\
2014-01-13T04:30:03 & -89 & 15 & E & N & N &  &  &  \\
2014-01-13T06:15:03 & 86 & 43 & NW & N & N &  &  &  \\
2014-01-16T11:30:03 & 88 & -28 & SW & Y & N & 7.70$\times$ $10^{14}$ & 268 & 3 \\
2014-01-21T07:52:03 & 90 & -02 & W & Y & N & 4.66$\times$ $10^{14}$ & 861 & -82 \\
2014-01-26T20:24:03 & -84 & -48 & SE & Y & Y & 1.60$\times$ $10^{15}$ & 450 & 63 \\
2014-01-29T00:00:03 & 61 & -31 & SW & N & N & no data & 254 & 4 \\
2014-01-30T13:14:03 & -53 & -20 & SE & Y & N & 3.02$\times$ $10^{15}$ & 1299 & -3 \\
2014-02-06T03:32:03 & -82 & -53 & SE & N & Y &  &  &  \\
2014-02-15T20:50:23 & 89 & 07 & W & N & N & no data & 207 & -47 \\
2014-02-17T04:30:35 & 89 & 03 & W & Y & Y & 1.24$\times$ $10^{15}$ & 354 & -16 \\
\hline\\
\end{tabular}
\end{table}
\begin{table}[htp]
\centering
\footnotesize\setlength{\tabcolsep}{4pt}
\begin{tabular}{lrrlccrrr}
\hline \hline
Start Date/Time & Long & Lat & Region & 3-part CME & AIA loops & Mass (g) & v (km.$s^{-1}$) & a (m.$s^{-2}$)
\\\hline
2014-02-18T14:00:03 & -69 & -72 & SE & Y & N & 4.56$\times$ $10^{14}$ & 671 & -48 \\
2014-02-18T21:30:03 & -61 & 28 & NE & Y & N & 4.70$\times$ $10^{14}$ & 428 & 34 \\
2014-02-25T00:40:03 & -89 & -12 & E & Y & Y & 1.45$\times$ $10^{14}$ & 1603 & -96 \\
2014-02-27T07:42:03 & -87 & -22 & SE & N & N &  &  &  \\
2014-02-28T19:24:03 & -90 & 00 & E & N & N & 3.68$\times$ $10^{14}$ & 447 & -20 \\
2014-03-01T04:20:03 & -89 & -11 & E & N & N & 6.47$\times$ $10^{14}$ & 564 & 13 \\
2014-03-02T09:00:03 & -90 & -00 & E & N & N &  &  &  \\
2014-03-04T20:00:03 & 51 & -50 & SW & N & N &  &  &  \\
2014-03-04T20:35:23 & -53 & 09 & E & Y & Y & 4.09$\times$ $10^{14}$ & 1082 & -136 \\
2014-03-10T15:36:03 & -45 & -20 & SE & Y & Y & 1.46$\times$ $10^{15}$ & 560 & 28 \\
2014-03-11T07:40:03 & 89 & -13 & W & N & N & 3.50$\times$ $10^{14}$ & 208 & 1 \\
2014-03-14T07:33:03 & -88 & -16 & E & Y & Y & 2.78$\times$ $10^{15}$ & 481 & 1 \\
2014-03-16T01:48:03 & -60 & -43 & SE & Y & Y & 7.08$\times$ $10^{14}$ & 599 & 17 \\
2014-03-18T13:00:03 & 84 & 41 & NW & N & N &  &  &  \\
2014-03-19T12:10:11 & 65 & 73 & NW & N & N &  &  &  \\
2014-03-26T04:50:23 & -82 & -51 & SE & Y & N & 5.77$\times$ $10^{14}$ & 579 & 22 \\
2014-03-26T07:00:03 & 86 & 27 & NW & N & Y &  &  &  \\
2014-03-27T12:30:35 & 65 & -26 & SW & N & N &  &  &  \\
2014-03-28T01:20:23 & 81 & -10 & SW & N & N &  &  &  \\
2014-03-29T01:20:03 & -46 & 27 & NE & N & N & no data & 572 & -30 \\
2014-03-30T10:24:03 & -86 & 29 & NE & Y & Y & 1.18$\times$ $10^{15}$ & 309 & 31 \\
2014-03-31T07:40:03 & 89 & -10 & W & Y & Y & 1.78$\times$ $10^{15}$ & 286 & -6 \\
2014-04-02T12:38:03 & -79 & 22 & NE & Y & Y & 3.06$\times$ $10^{15}$ & 1322 & 18 \\
2014-04-03T13:24:03 & 72 & -46 & SW & Y & Y & 1.50$\times$ $10^{15}$ & 440 & -6 \\
2014-04-03T14:10:03 & 89 & 08 & W & N & N &  &  &  \\
2014-04-03T18:00:03 & 72 & -71 & SW & Y & Y & 1.50$\times$ $10^{15}$ & 171 & 21 \\
2014-04-04T13:30:03 & -36 & 24 & NE & Y & N & 8.67$\times$ $10^{14}$ & 800 & -29 \\
2014-04-04T23:10:03 & -86 & -34 & SE & Y & N & 8.06$\times$ $10^{14}$ & 598 & -5 \\
2014-04-06T09:08:03 & 88 & 15 & NW & Y & N & 5.80$\times$ $10^{14}$ & 421 & 8 \\
2014-04-06T14:00:03 & 88 & 19 & NW & N & N &  &  &  \\
2014-04-07T20:30:03 & 86 & -34 & SW & Y & Y & 1.54$\times$ $10^{15}$ & 428 & 15 \\
2014-04-09T14:06:03 & 90 & 01 & W & N & N &  &  &  \\
2014-04-12T07:00:03 & -89 & -11 & E & Y & Y & 4.71$\times$ $10^{15}$ & 911 & -188 \\
2014-04-14T21:10:03 & -90 & 01 & E & N & N &  &  &  \\
2014-04-19T06:30:35 & 70 & 11 & W & N & N &  &  &  \\
2014-04-21T00:15:35 & 87 & -32 & SW & N & N &  &  &  \\
2014-04-21T10:30:35 & 88 & 23 & NW & N & N & 6.88$\times$ $10^{14}$ & 211 & -1 \\
2014-04-22T04:44:03 & -60 & -81 & S & N & N & 7.03$\times$ $10^{13}$ & 356 & -3 \\
2014-04-22T08:00:03 & -48 & 25 & NE & Y & N & 4.86$\times$ $10^{14}$ & 472 & 12 \\
2014-04-23T06:46:03 & -84 & -54 & SE & N & N &  &  &  \\
2014-04-29T04:36:03 & -44 & -60 & S & N & N &  &  &  \\
2014-04-29T18:04:03 & 87 & -36 & SW & Y & N & 4.08$\times$ $10^{14}$ & 182 & 3 \\
2014-04-30T07:12:03 & -51 & 40 & NE & N & N &  &  &  \\
2014-04-30T23:00:03 & 77 & 71 & N & N & N &  &  &  \\
2014-05-01T20:20:03 & -88 & 28 & NE & N & N &  &  &  \\
2014-05-03T07:30:03 & -67 & 12 & E & N & N &  &  &  \\
2014-05-04T14:16:03 & 88 & 21 & NW & Y & N & 1.07$\times$ $10^{15}$ & 263 & 4 \\
2014-05-05T14:00:03 & 89 & 07 & W & N & N &  &  &  \\
2014-05-07T12:00:03 & 89 & -17 & W & Y & N & 1.07$\times$ $10^{15}$ & 1101 & -105 \\
2014-05-08T02:00:03 & -89 & -26 & SE & N & N & 2.69$\times$ $10^{14}$ & 618 & -68 \\
2014-05-09T01:00:03 & 77 & -16 & W & Y & N & 1.47$\times$ $10^{15}$ & 851 & no data \\
2014-05-10T22:50:03 & -88 & 34 & NE & N & N &  &  &  \\
\hline\\
\end{tabular}
\end{table}
\begin{table}[tp]
\centering
\footnotesize\setlength{\tabcolsep}{4pt}
\begin{tabular}{lrrlccrrr}
\hline \hline
Start Date/Time & Long & Lat & Region & 3-part CME & AIA loops & Mass (g) & v (km.$s^{-1}$) & a (m.$s^{-2}$)
\\\hline
2014-05-15T17:20:03 & 85 & 62 & NW & N & N &  &  &  \\
2014-05-15T19:40:03 & -88 & -38 & SE & Y & Y & 2.17$\times$ $10^{15}$ & 620 & 24 \\
2014-05-23T04:10:35 & -89 & -35 & SE & N & N & 2.70$\times$ $10^{14}$ & 462 & 9 \\
2014-05-26T09:35:03 & -67 & 34 & NE & N & N & 1.21$\times$ $10^{15}$ & 395 & 37 \\
2014-05-27T13:00:03 & 89 & -34 & SW & N & Y & 1.13$\times$ $10^{14}$ & 213 & -8 \\
2014-05-28T10:50:23 & 89 & 13 & W & N & N &  &  &  \\
2014-05-28T12:00:03 & 89 & -48 & SW & N & N & 6.46$\times$ $10^{14}$ & 554 & 107 \\
2014-05-30T05:00:03 & 73 & 86 & N & N & N &  &  &  \\
2014-05-30T23:30:35 & 90 & -35 & SW & N & N & 2.70$\times$ $10^{14}$ & 209 & 11 \\
2014-06-04T10:16:03 & -52 & -36 & SE & Y & Y & 1.41$\times$ $10^{15}$ & 226 & 42 \\
2014-06-08T22:59:59 & -63 & -18 & E & Y & N & 1.02$\times$ $10^{15}$ & no data & no data \\
2014-06-13T00:00:03 & -90 & 26 & NE & N & N & no data & no data & no data \\
2014-06-13T00:00:03 & -89 & -33 & SE & N & N & no data & no data & no data \\
2014-06-14T19:00:33 & -90 & -12 & E & Y & Y & 1.35x15 & 756 & 135 \\
2014-06-15T12:16:03 & -89 & -31 & SE & Y & N & 5.99$\times$ $10^{14}$ & 531 & 51 \\
2014-06-16T08:12:03 & 89 & 38 & NW & N & N & no data & 267 & 17 \\
2014-06-30T16:40:11 & -89 & -21 & SE & Y & N & 2.46$\times$ $10^{15}$ & 653 & 136 \\
2014-06-30T17:00:03 & 89 & -20 & W & Y & N & no data & no data & no data \\
\hline
\end{tabular}
\end{table}

\end{document}